\documentstyle[12pt]{article}

\begin{document}

\thispagestyle{empty}
\setcounter{page}{0}

\hfill\vbox{
  \hbox{OHSTPY-HEP-T-96-019}
  \hbox{hep-ph/9612291}
  \hbox{October 1996}
}\par

\vspace*{\stretch{.5}}

\centerline{\bf PERTURBATIVE QCD AT HIGH 
TEMPERATURE\footnote{Published in {\em Int.J.Mod.Phys.\/} A {\bf 12},
1431 (1997)}}

\vspace*{\stretch{.25}}

\centerline{\footnotesize AGUSTIN NIETO}
\vspace*{0.015truein}
\centerline{\footnotesize\it Department of Physics, The Ohio State 
University}
\centerline{\footnotesize\it Columbus, OH 43210, USA}
\vspace*{0.225truein}

\vspace*{\stretch{1}}

\begin{abstract}
Recent developments of perturbation theory at finite
temperature based on effective field theory methods are
reviewed. These methods allow the contributions from the
different scales to be separated and the perturbative series
to be reorganized.  The construction of the effective field
theory is shown in detail for $\phi^4$ theory and QCD. It is
applied to the evaluation of the free energy of QCD at order
$g^5$ and the calculation of the $g^6$ term is outlined.
Implications for the application of perturbative QCD to the
quark-gluon plasma are also discussed.
\end{abstract}

\vspace*{\stretch{1}}

\newpage

\setcounter{footnote}{0}
\renewcommand{\thefootnote}{\alph{footnote}}

\section{Introduction}
\noindent
In this paper I review some of the recent developments in
perturbative field theory at finite temperature that have
come about by using effective field theory methods.  In particular,
we will be interested in the behavior of hadronic matter at
high temperature. This means that the temperature is assumed
to be much larger than the mass of the particles involved
($T\gg m$).  Another assumption is that the gauge coupling
constant $g$ is small ($g\ll 1$); this allows us to define
three energy scales that satisfy: $T\gg gT\gg
g^2T$. Hadronic matter is expected to undergo a phase
transition when $T\sim T_c\simeq 200$~MeV between a low
temperature phase in which it is confined in the form of
hadrons and a high temperature phase in which quarks and
gluons are deconfined.  The latter phase, known as {\em
Quark-Gluon Plasma} (QGP), can be studied by perturbative
methods since the strong coupling constant $\alpha_s(T)$ is
expected to be small at high temperature. Later on, we will
check the validity of this statement; it happens that at
temperatures of a few times $T_c$, the strong coupling
constant is not small enough to give rise to a convergent
series in the case of the free energy.

A field theory at high temperature may be described by a
lagrangian in which the modes with zero Matsubara frequency
have been decoupled according to the Appelquist-Carazzone
theorem~\cite{Appelquist-Carazzone}. Since the resulting
theory is 3-dimensional, this method is know as dimensional
reduction~\cite{appelquist-pisarski,nadkarni-1,gpy}.  An
efficient approach consists of interpreting the
dimensionally reduced theory as an effective field
theory~\cite{georgi} whose parameters are to be computed as a
perturbative expansion in the coupling constant of the
original theory.  This idea is very attractive when studying
theories with fermions on the lattice: the dimensionally
reduced theory does not contain fermions because they have
been integrated out completely and this may simplify
computer simulations~\cite{i-r-k}. The parameters of the
effective theory can be determined by analytic methods.

Effective field theory methods have been applied to
different problems.  Braaten~\cite{solution} has resolved a
longstanding problem involving the breakdown of the
perturbation expansion for the free energy of QCD. Braaten
and Nieto have determined the asymptotic behavior of the
correlator of Polyakov loops~\cite{polyakov}. They have also
used this method to carry out explicit calculations in
$\phi^4$~\cite{eft} and in QCD~\cite{fQCD,bn5}.  The problem
of the finite $T$ electroweak phase transition has
essentially been solved in~\cite{f-k-r-s}; this work has
recently been reviewed by Shaposhnikov in an
article~\cite{shap} that is complementary to this
review. Andersen~\cite{andersen} has analytically computed
the free energy of QED up to order $e^5$ and the electric
screening mass of scalar QED up to order $e^4$. Arnold and
Yaffe~\cite{ay} have proposed a rigorous nonperturbative
definition of the Debye screening mass in nonabelian gauge
theories. Recently, Karsch~{\em et~al.\/}~\cite{Karsch} have
computed the nonperturbative contribution to the $g^6$-order
term of the free energy of QCD~\cite{solution,bn5} by using
lattice simulations. Also, the effective field theory
approach has been applied to the study of the Minimal
Supersymmetric Standard Model~\cite{MSSM}.

The effective field theory approach is now a
well-established perturbative method to study field theories
at finite temperature. It is a useful tool for organizing
calculations of high-order terms in the perturbation
expansion and it is especially powerful in dealing with
nonabelian gauge theories. It is also an important
conceptual tool that explicitly separates the different
energy scales.  In this review I present some of the results
of its application to QCD. There are two conceptual steps:
first, the effective field theory is defined by computing
its parameters by matching with the original theory; second,
the effective theory is used to study problems concerning
the high temperature limit of the full theory.

We will use dimensional reduction to construct an effective
field theory of QCD whose parameters encode the physics from
the scale $T$; this theory is called {\em Electrostatic QCD}
(EQCD), because it consists of pure Yang-Mills in three
dimensions plus a scalar field in the adjoint representation
which is related to the static mode of the original
chromelectric field.  Since, the effective mass of the
scalar is of order $gT$ and magnetostatic fields remain
massless, we can use the decoupling theorem to construct
another effective theory in which the scalar field is
decoupled. We will end up with a pure Yang-Mills theory in
three dimensions that is called {\em Magnetostatic QCD}
(MQCD) because it is only made out of fields related to the
magnetostatic modes of QCD; its parameters encode the
physics from the scale $gT$.  In both EQCD and MQCD, the
effective lagrangian includes an infinite series of
nonrenormalizable terms.  One of the technically more
involved problems in constructing the effective field theory
is the evaluation of Feynman diagrams. In what follows I
will not give details of these calculations, I will just
state the results of the integrals that are needed and give
references, which are basically the appendices
in~\cite{eft,bn5}. In all other respects I will try to show
the construction of the effective theory in detail.

The next section reviews some concepts that are familiar in
the context of quantum field theory at finite temperature
(more information can be found in the
textbooks~\cite{kapbook,lebellac}), introduces the effective
field theory approach for $\phi^4$ theory, and shows a
complete calculation using its effective field theory. In
Sect.~\ref{separation}, the contributions to the QCD free
energy from the relevant scales are
separated. Sect.~\ref{parameters} is devoted to the
construction of EQCD. The free energy of QCD is explicitly
computed in Sect.~\ref{fenQCD} up to order $g^5$; also the
result at order $g^6$ is analyzed. In Sect.~\ref{conver} we
will study the convergence of the perturbative series for
the free energy.  Conclusions and final comments are given
in Sect.~\ref{conclusions}.

\section{Field Theory at Finite Temperature}
\label{ftft}

The static properties of a system of particles in thermal
equilibrium at temperature $T$ are obtained from the free
energy density
\begin{equation}
  F = -{T\over V} \log{\cal Z}  \, ,
\end{equation}
where ${\cal Z}$ is the partition function. All through this
paper, we will use the imaginary-time formalism which is more
suitable than the real-time formalism for studying static
properties. The partition function of a theory for a field
$\Phi$ is, in general, of the form
\begin{equation}
  {\cal Z}(\beta) = \int {\cal D}\Phi({\bf x},\tau)\exp
  \left\{
    -\int_0^\beta d\tau\int d{\bf x}\, {\cal L}(\Phi)
  \right\}\, ,
\end{equation}
where $\tau$ is the Euclidean time and $\beta=1/T$. Also,
$\Phi$ satisfies periodic boundary conditions in $\tau$ if
it represents a bosonic field or anti-periodic boundary
conditions if it represents a fermionic field:
\begin{equation}\label{bcond}
  \Phi({\bf x},0) = \pm \Phi({\bf x},\beta)\, .
\end{equation}
These conditions allow us to expand $\Phi$ in its
Fourier modes:
\begin{equation}\label{modes}
  \Phi({\bf x},\tau) =
    T \sum_{n}\phi_n({\bf x})e^{i\omega_n\tau}\, .
\end{equation}
The Matsubara frequency of the $n$th mode is
$\omega_n=(2n)\pi T$ for bosons and $\omega_n=(2n+1)\pi T$
for fermions. The term in~(\ref{modes}) with $\omega_n=0$ is
independent of the Euclidean time $\tau$. The mode with
this zero Matsubara frequency is called {\em static\/}; the
rest of the modes involve $\tau$ and are called {\em
nonstatic\/}.

The Feynman rules are the same as in the Euclidean field
theory, except that, at finite temperature, the time-like
component of the momentum is discrete: $K^\mu=(\omega_n,{\bf
k})$.  Consequently, the loop integration over such a
component is replaced by a sum and the rule for loops is
\begin{equation}
\hbox{$\sum$}\!\!\!\!\!\!\int_K \equiv
\left({e^\gamma \mu^2 \over 4 \pi}\right)^\epsilon\;
T \sum_{K_0 = \omega_n} \:
  \int {d^{3-2\epsilon}k \over (2 \pi)^{3-2\epsilon}}
\,,
\end{equation}
where $3-2\epsilon$ is the dimension of the space and $\mu$
is an arbitrary renormalization scale. The factor
$(e^\gamma/4\pi)^\epsilon$ is introduced so that, after
minimal subtraction of the poles in $\epsilon$ due to
ultraviolet divergences, $\mu$ coincides with the
renormalization scale in the $\overline{\rm MS}$
renormalization scheme.

For example, in the $\phi^4$ theory with a massless scalar field and
interaction $g^2\Phi^4$, described by the lagrangian
\begin{equation}\label{phi4}
{\cal L} = {1\over 2} (\partial\Phi)^2 + {g^2\over 4!}\Phi^4 \, ,
\end{equation}
the leading-order contribution to the self-energy for a particle
of momentum $P^\mu=(E_n,{\bf p})$ is given by
the tadpole diagram in Fig.~\ref{tadpole}$(a)$:
\begin{equation}\label{pione}
\Pi^{(1)}(P) =
  {g^2\over 2} \hbox{$\sum$}\!\!\!\!\!\!\int_K
  {1\over K^2}
  = {g^2 T^2 \over 24}\, ,
\end{equation}
where $K^2 = \omega_n^2 + k^2$. 

If we try to compute the next-order correction to the
self-energy which is given by the Fig.~\ref{tadpole}$(b)$,
we will find out that it is infrared divergent. The same is
true for the rest of the diagrams shown in
Fig.~\ref{tadpole}. This {\em mild} breakdown of the
perturbative expansion may be cured by resumming the
geometric series of diagrams with one-loop-tadpole
insertions like those in the Fig.~\ref{tadpole}. The
result is finite and gives the self-energy at
next-to-leading order in $g$:
\begin{equation}
  \Pi =
  {g^2\over 2} \hbox{$\sum$}\!\!\!\!\!\!\int_K
  {1\over K^2 + \Pi^{(1)}} =
  {g^2 T^2 \over 24}
    \left( 1 - {\sqrt{6}\over 4\pi}g + {\cal O}(g^2) \right)\, .
\end{equation}
The sum-integral may be found in~\cite{arnold-zhai}.  We see
that the perturbative series is not analytic in the coupling
constant $g^2$, but in $g$. This is a
general feature of the resummation of diagrams with
insertions of the leading-order contribution of the
self-energy.

Ultraviolet divergences are not a special issue at finite
temperature. The very same counterterms that would
regularize the theory at zero temperature regularize the
finite temperature theory. This can be understood by
realizing that a theory at finite temperature is defined in
a 4-dimensional Euclidean space whose time component is
compactified into a circle of circumference $1/T$. The
short-distance behavior is not modified by this sort of
compactification.

\subsection{Dimensional Reduction}

Let us consider the Appelquist-Carazzone decoupling theorem
for a field theory at zero temperature. It states that for a
renormalizable field theory with heavy fields (let us say,
with mass $\sim M$) and light fields (with mass $\sim m\ll
M$), Green's functions with typical momentum scale $p\ll M$
and that only involve light fields in the external legs may
be computed without considering heavy field loops.  Up to
corrections suppressed by powers of $p/M$ and $m/M$, such
Green's functions are described by the original lagrangian
with the heavy fields removed, but with modified values for
the coupling constants of the light fields. In this sense,
the heavy fields decouple from the light fields.

The partition function of a field theory at finite $T$ may
be written in terms of the Fourier modes $\phi_n({\bf x})$,
defined in~(\ref{modes}), instead of $\Phi({\bf x},\tau)$:
\begin{equation}\label{imodes}
  {\cal Z} =
  \int{\cal D}\phi_0({\bf x}){\cal D}\phi_n({\bf x})\exp
  \left\{
    -\int_0^\beta d\tau\int d{\bf x}\,
    {\cal L}(\phi_0,\phi_n,\tau)
  \right\}\, .
\end{equation}
A propagator of the form $1/(\omega_n^2+k^2)$ can be
associated with each mode and the corresponding Matsubara
frequency represents its mass. In this context, all of the
fermionic modes and the nonstatic bosonic modes have a mass
of order $T$, while the static bosonic modes are massless.

In the limit of high temperature, nonstatic modes have a
large mass of order $T$ when compared with static modes, which are
massless; therefore, the decoupling theorem suggests that
they decouple. What remains is an effective theory of the
static modes whose partition function is of the form
\begin{equation}\label{effz}
  {\cal Z} =
  \int{\cal D}\phi({\bf x})\exp
  \left\{
    -\int d{\bf x}\,
    {\cal L}_{\rm eff}(\phi)
  \right\}\, ,
\end{equation}
where $\phi({\bf x})\equiv\sqrt{T}\phi_0({\bf x})$ and
${\cal L}_{\rm eff}$ is the effective lagrangian.  The
effective theory is defined in a 3 dimensional space; this
is why this procedure is known as {\em dimensional
reduction}. Note that all fermionic modes have masses of
order $T$ and therefore are integrated out. There are no
effective fields associated with them. Their effects are all
incorporated into the effective theory through the effective
coupling constants for the bosonic modes.

There is an important difference between the decoupling
theorem applied to heavy fields at zero temperature and to
dimensional reduction at high temperature~\cite{gpy}. For
heavy fields at zero temperature, the effective theory is
usually taken to be renormalizable. It reproduces Green's
functions of the full theory to all orders in the coupling
constant $g$, up to corrections that fall like powers of
$p/M$ and $m/M$. It is possible to reproduce these
corrections as well but only at the cost of introducing
nonrenormalizable terms into the effective lagrangian.  As
we have seen for the $g^2\Phi^4$ theory at high temperature,
perturbative corrections generate a mass $m$ for the field
of order $gT$, while the nonstatic field has a mass $M$ of
order $T$. The renormalizable effective theory provided by
the decoupling theorem reproduces Green's functions only up
to errors that fall like powers of $m/M$, but in this case
the errors correspond to powers of $g$. Thus the
renormalizable theory does not reproduce the full theory to
all orders in $g$.  Landsman~\cite{landsman} has concluded
that the validity of the decoupling theorem at finite $T$
depends on the theory under study and fails for QCD or
$\phi^4$ theory. This is true only if dimensional reduction
is taken in the restricted sense of reducing to a
renormalizable effective theory.  As in the case of heavy
fields at zero temperature, corrections that fall like
powers of $m/M$ can be reproduced by including
nonrenormalizable terms in the effective theory.

The effective lagrangian therefore has an infinite number of
interaction terms; only a few of them are renormalizable,
while the rest are non-renormalizable. Each term has a
parameter that plays the role of an effective coupling
constant. These parameters are not arbitrary coefficients
but instead are completely determined by the condition that
the effective theory matches the full theory at low
momentum. These effective parameters can be calculated in
powers of the coupling constant $g$ of the full
theory. Also, in general, the effective parameters depend on
an ultraviolet cutoff that cancels the ultraviolet cutoff
dependence of the loop integrals in the effective theory.
In calculations of a physical quantity to a given order in
$g$, only a finite number of these parameters will enter. At
low orders in $g$, it is sometimes possible to truncate the
effective theory to the renormalizable terms or even to the
super-renormalizable terms.

\subsection{A nontrivial example: the screening mass
of $\phi^4$ theory}
\label{ane}

In this subsection we will explicitly construct the
effective field theory of a massless field with
self-interaction $g^2\Phi^4$ described by the
lagrangian~(\ref{phi4}) and apply it to evaluate the
screening mass. 

In QED, the electric screening mass describes the asymptotic
behavior of the potential between two static charges at
large distances. The potential is
\begin{equation}\label{potmel-1}
  V(R)=\int {d^3k\over (2\pi)^3}\;
    {e^{-i {\bf k}\cdot{\bf R}}\over k^2 +
      \Pi_{00}(0,{\bf k})} \, ,
\end{equation}
where $\Pi_{\mu\nu}$ is the photon self-energy. The 
asymptotic behavior of the potential is 
\begin{equation}\label{potmel-2}
  V(R\rightarrow\infty)\sim
    {e^{-m_{\rm el}R}\over R} \, .
\end{equation}
We observe that the potential at large distances is not a
Coulomb potential ($1/r$) but a Yukawa potential
($e^{-\alpha r}/r$), so that it goes to zero very rapidly at
distances larger that $1/m_{\rm el}$.  That is why $m_{\rm
el}$ is called {\em screening mass}.  The asymptotic
behavior of the Fourier transform in~(\ref{potmel-1}) is
determined by the singularity of the integrand closest to
the origin, which in this case is a pole in the propagator
$1/(k^2 + \Pi_{00}(0,{\bf k}))$; therefore, the electric
screening mass satisfies~\cite{rebhan}:
\begin{equation}\label{polesm}
k^2 \;+\; \Pi_{00}(0,{\bf k}) \;=\; 0
\qquad \mbox{at $k^2=-m_{\rm el}^2$} \, .
\end{equation}
Analogously, the screening mass $m_s$ for the $\Phi^4$
theory is given by the location of the pole of the static
($\omega=0$) propagator:
\begin{equation}\label{smass}
  k^2 \;+\; \Pi(0,{\bf k}) \;=\; 0
  \qquad \mbox{at $k^2=-m_{\rm s}^2$} \, ,
\end{equation}
where $\Pi(\omega,{\bf k})$ is the self-energy of $\Phi$.

The effective lagrangian has to be compatible with the
symmetries of the original theory; {\em i.e.\/}, it has to
be symmetric under the exchange $\phi\rightarrow-\phi$.
Then, the most general lagrangian is of the form
\begin{equation}\label{eff4}
  {\cal L}_{\rm eff} =
  {1 \over 2} \left( \makebox{\boldmath $\nabla$} \phi \right)^2
  \;+\; {1 \over 2} m^2 \; \phi^2
  \;+\; {1 \over 4!} \lambda \; \phi^4
  \;+\; {1 \over 6!} \lambda_6 \; \phi^6
  \;+\; \delta {\cal L} \, ,
\end{equation}
where $m^2$, $\lambda$, and $\lambda_6$ are coupling
constants of the effective theory.  $\delta {\cal L}$
represents an infinite series of nonrenormalizable terms,
each of them with a corresponding effective coupling
constant. At leading order, any effective parameter will be
proportional to $T$ raised to the power required by
dimensional analysis; the leading power of $g^2$ is
determined by identifying the lowest order diagram that
contributes to that parameter. The leading-order
contributions to $m^2$, $\lambda$, and $\lambda_6$ are shown
in Figs.~\ref{poco} $(a)$, $(b)$, and $(c)$,
respectively. By counting the powers of the coupling
constant in the diagrams and using dimensional analysis to
determine the powers of $T$, their magnitudes must be
\begin{eqnarray}
  m^2        & \sim & g^2 T^2 \, ,  \\
  \lambda    & \sim & g^2 T   \, ,  \\
  \lambda_6  & \sim & g^6     \, .
\end{eqnarray}
If we are interested in calculations up to order
$g^4$, we can drop the $\phi^6$-term in ${\cal L}_{\rm
eff}$, since it only contributes at higher order. Similarly,
the non-renormalizable terms included in $\delta{\cal L}$
can also be neglected. What remains is a
super-renormalizable effective theory. 

The effective coupling $\lambda$ is computed by matching the
full theory and the effective theory. Let us consider the
action of the full theory defined by the
lagrangian~(\ref{phi4})
\begin{equation}
  S_{\rm full} = \int_0^{1/T} d\tau \int d^3x\; \left[ 
    {1\over 2}(\partial\Phi)^2 + {g^2\over 4!}\Phi^4  \right] \, ,
\end{equation}
Now, we use the expansion of $\Phi$ in terms of its Fourier
modes~(\ref{modes}), 
\begin{equation}
  \Phi({\bf x},\tau) =
    T \sum_{n}\phi_n({\bf x})e^{i\omega_n\tau}\, ,
\end{equation}
and write $S_{\rm full}$ as the sum of two terms; the first
of them only depends on the static mode of the field and
the second term also depends on the nonstatic modes. After
integrating over $\tau$, we get
\begin{equation}\label{sfull}
  S_{\rm full} = \int d^3x\; \left[
    {1\over 2} T ({\bf \nabla}\phi_0)^2 + 
    {1\over 4!} g^2 T^3 \phi_0^4  \right]+ 
    \widetilde{S}_{\rm full}[\phi_0,\phi_n]  \, .
\end{equation}
Since the first term does not depend on the nonstatic modes,
it does not change after integrating over the nonstatic
modes. As a consequence it can be compared directly with the
action of the effective theory. After rescaling
$\phi\rightarrow\sqrt{T}\phi_0$ in the effective
lagrangian~(\ref{eff4}) so that the kinetic terms match, we
obtain
\begin{equation}\label{seff}
  S_{\rm eff} = \int d^3x\; \left[
    {1\over 2} T ({\bf \nabla}\phi_0)^2 + 
    {1\over 4!} \lambda T^2 \phi_0^4 \right]  \, .
\end{equation}
Comparing~(\ref{sfull}) and~(\ref{seff}), we conclude that
\begin{equation}\label{efflam}
  \lambda = g^2 T
\end{equation}
We have ignored the integration of $\widetilde{S}_{\rm
full}$ over the nonstatic fields $\phi_n$ because we only
want to obtain $\lambda$ at leading order.  This integration
contributes to $\lambda$ starting at next-to-leading-order
and generates the contributions to the other parameters of
the effective lagrangian~(\ref{eff4}) such as $m^2$,
$\lambda_6$, etc.

The actual evaluation of the effective parameters is, in
general, performed by matching the full and effective
theories in the region where they describe the same
physics. This region corresponds to large distances $R\gg
1/T$; therefore, we have to match physical quantities whose
Feynman diagrams have a typical external momentum much
smaller than $T$ ($p_{\rm ext} \ll T$).  In particular,
$m^2$ is determined by matching the electric screening mass
$m_s$. In the effective theory, the screening mass is
defined by
\begin{equation}\label{emass}
  k^2 \;+\; m^2 \;+\; \Pi_{\rm eff}({\bf k}) \;=\; 0
  \qquad \mbox{at $k^2=-m_{\rm s}^2$} \, ,
\end{equation}
where $\Pi_{\rm eff}({\bf k})$ is the self-energy of $\phi$.
We could compute $m_s$ in both the full theory and the
effective theory and then match them to determine $m^2$. In
the full theory we would have to resum infinite series
of diagrams such as those described at the beginning of
Sect.~\ref{ftft}.

There is an alternative method that greatly reduces the
effort required to determine the parameters in the effective
theory. It involves introducing an infrared cutoff
$\Lambda_{IR}$ satisfying $gT\ll\Lambda_{IR}\ll T$ on the
internal momenta $p_{\rm int}$ of diagrams contributing to
the physical quantity that we want to compute.  In the full
theory the infrared cutoff should satisfy
$gT\ll\Lambda_{IR}\ll T$ and in the effective theory
$m\ll\Lambda_{IR}\ll\Lambda$, where $\Lambda$ is the
ultraviolet cutoff of the effective theory.  In the full
theory with an infrared cutoff $p_{\rm int}>\Lambda_{IR}$,
there are no complications from infrared divergences since
$\Lambda_{IR}\gg gT$. We can therefore avoid resummations
and compute using the {\em strict perturbation expansion} in
powers of $g^2$. In the effective theory, with an infrared
cutoff $p_{\rm int}>\Lambda_{IR}$, calculations can also be
simplified. Since the leading order contribution to $m^2$ is
of order $g^2T^2$ and $\Lambda_{IR}\gg gT$, we can treat
$m^2$ as a perturbation and take the propagator of $\phi$ to
be simply $1/k^2$. In what follows I will refer to the
resulting perturbative expansion as the {\em strict
perturbation expansion} of the effective theory.  By
construction, our effective theory is equivalent to the full
theory at low momentum. Therefore, the infrared cutoff
modifies the full theory and the effective theory in
precisely the same way. We can therefore match the strict
perturbation expansions of the full and effective theories
to extract the effective parameters.

Note that we use the strict perturbation expansion only as a
device to determine the effective parameters. To actually
calculate physical quantities we must remove the infrared
cutoff. In the effective theory, this will require using
$1/(k^2+m^2)$ as the propagator of the field $\phi$ instead
of $1/k^2$.

Although we have used a momentum cutoff for illustration, we
can determine the effective parameters by matching the
strict perturbation expansions of the full and effective
theories with any infrared regulator.  As we will see, a
particularly convenient choice for the infrared cutoff is
dimensional regularization. While this method does not
explicitly cut off the low momentum region from internal
loops, it still allows the consistent calculation of a
strict perturbation expansion for both the full theory and
the effective theory.

We now proceed to calculate the mass parameter $m^2$ in the
effective lagrangian~(\ref{eff4}) for the $\Phi^4$ theory by
matching calculations of the screening mass defined
by~(\ref{smass}) in the full theory and by~(\ref{emass}) in
the effective theory. The diagrams in the full theory that
contribute to $\Pi(0,{\bf k})$ are the ones shown in
Fig.~\ref{oloop}. Now, if the $g^{2n}$-order contribution to
$\Pi(k^2)$ is called $\Pi^{(n)}(k^2)$, the
definition~(\ref{smass}) tells us that:
\begin{equation}\label{mst2}
  m_{\rm s}^2 = \Pi(k^2=-m_{\rm s}^2)
  = \Pi^{(1)}(-m_{\rm s}^2) + \Pi^{(2)}(-m_{\rm s}^2) + \cdots \,.
\end{equation}
The first term $\Pi^{(1)}(-m_{\rm s}^2)$ is given explicitly
in~(\ref{pione}). It is independent of its argument and of
order $g^2T^2$. Therefore, at leading order $m_s\sim
g^2T^2$.  Since the argument of $\Pi^{(n)}(-m_s^2)$ is of
order $g^2T^2$, the strict perturbation expansion for the
right side of~(\ref{mst2}) is obtained by making a Taylor
expansion around 0.  $\Pi^{(2)}(-m_s^2)$ is already of order
$g^4$, we can set its argument to zero up to corrections of
order $g^4$. We conclude that
\begin{equation}\label{msone}
  m_{\rm s}^2 \approx 
  {Z^2_g g^2 \over 2}\; \hbox{$\sum$}\!\!\!\!\!\!\int_P {1 \over P^2}
  \;-\; {g^4 \over 4}\; \hbox{$\sum$}\!\!\!\!\!\!\int_P {1 \over P^2}
  \hbox{$\sum$}\!\!\!\!\!\!\int_P {1 \over (P^2)^2}
  \;-\; {g^4 \over 6}\; \hbox{$\sum$}\!\!\!\!\!\!\int_{PQ}
  {1 \over P^2 Q^2 (P+Q)^2} \, .
\end{equation}
Here and below, the symbol ``$\approx$'' denotes an equality
that holds only in the strict perturbation expansion.  
To the order required, renormalization of the coupling constant
in the $\overline{\rm MS}$ scheme is accomplished by the substitution
\begin{equation}
Z_g \;=\;
1 \;+\; {3 \over 4 \epsilon} \left({g\over 4\pi}\right)^2 \, .
\end{equation}
The integrals in~(\ref{msone}) are evaluated in Ref.~\cite{eft}
and the final result is
\begin{equation}\label{ms2x}
  m_s^2 \approx
  {1\over 24}g^2 T^2\left\{
    1 + \left[{1\over\epsilon}
    + \log{\Lambda\over 4\pi T} + 2 - \gamma
    + 2{\zeta'(-1)\over\zeta(-1)}\right]
      \left({g\over 4\pi}\right)^2 
    \right\}  \, ,
\end{equation}
where $\Lambda$ is the mass scale introduced by dimensional
regularization.  

The diagrams in the effective theory that contribute to
$\Pi_{\rm eff}({\bf k})$ are shown in Fig.~\ref{oloop}
(where, the propagators and vertices are now those of the
effective theory) and in Fig.~\ref{ctdia}. After expanding
around $k^2=0$, there is no mass scale in the integrals and
the loop diagrams vanish in dimensional regularization,
\begin{equation}
\Pi_{\rm eff}(0) = \delta m^2 \, .
\end{equation}
Consequently, from the definition~(\ref{emass}),
\begin{equation}\label{mstwo}
m_{\rm s}^2 \approx m^2 + \delta m^2\, .
\end{equation}

We obtain $m^2$ by matching~(\ref{ms2x}) and~(\ref{mstwo}).
$\delta m^2$ in~(\ref{mstwo}) is the mass counterterm that
contains the poles in $\epsilon$ that are associated with the
mass renormalization. Therefore, it is determined to be
\begin{equation}
  \delta m^2 = {1\over 24} g^2 T^2 
    \left(g\over 4\pi\right)^2 {1\over\epsilon}
\end{equation}
The $\Lambda$-dependence from the logarithm in~(\ref{msone})
is partially cancelled by the $\Lambda$-dependence of the
running coupling constant.  The renormalization group
equation for the coupling constant,
\begin{equation}
  \mu{d\over d\mu} \left({g\over 4\pi}\right)^2
  = 3 \left({g\over 4\pi}\right)^4 \, ,
\end{equation}
gives
\begin{equation}
  g^2(\Lambda) = g^2(\mu) \left[
    1 - 3 \left({g\over 4\pi}\right)^2 \log{\mu\over\Lambda}\right]\, ,
\end{equation}
which can be used to shift the renormalization scale
$\Lambda$ to an arbitrary scale $\mu$. We conclude
\begin{equation}\label{mlambda}
  m^2(\Lambda) =
  {1\over 24}g^2(\mu)T^2\left\{
    1 + \left[- 3 \log{\mu\over 4\pi T}
    + 4 \log{\Lambda\over 4\pi T} + 2 - \gamma
    + 2{\zeta'(-1)\over\zeta(-1)}\right]
      \left({g\over 4\pi}\right)^2 \right\} \, .
\end{equation}

Note that $m^2(\Lambda)$ depends logarithmically on the
ultraviolet cutoff $\Lambda$ at order $g^4$; this dependence
is necessary to cancel logarithmic ultraviolet divergences
from loop integrals in the effective theory. One should not
confuse this ultraviolet cutoff $\Lambda$ of the effective
theory with the infrared cutoff $\Lambda_{IR}$.  The latter
is removed in the matching process. We can explicitly see
this while using dimensional regularization by introducing
two different regularization scales $\Lambda$ and
$\Lambda_{IR}$ to regularize the ultraviolet and infrared
divergences respectively.  The scale $\Lambda$
in~(\ref{ms2x}) should be identified with the infrared scale
$\Lambda_{IR}$.  The expression~(\ref{mstwo}) for the
screening mass in the effective theory is then modified to
\begin{equation}
  m_s^2\approx
  m^2 (\Lambda) + {1\over 24} g^2T^2
    \left[ {1\over\epsilon} -
    \left({g\over 4\pi}\right)^2 \log{\Lambda\over\Lambda_{IR}} 
    \right] \, .
\end{equation}
Matching both results the dependence on $\Lambda_{IR}$
cancels and we recover~(\ref{mlambda}) for $m^2(\Lambda)$.

Now, having determined $m^2$, we can use the effective
theory to compute the screening mass.  Remember that now,
the $\phi$ field propagator is $1/(k^2+m^2)$, instead of
$1/k^2$.  Considering~(\ref{emass}), $\Pi_{\rm eff}$ is
given by the diagrams shown in Figs.~\ref{oloop}
and~\ref{ctdia} evaluated at the point $k=im$. We
obtain~\cite{eft}
\begin{equation}
  m_s^2\;=\;m^2(\Lambda) \Bigg\{ 1 \;-\; 2 \; {\lambda\over 16 \pi m}
\;-\; {2 \over 3} \left[4 \log{\Lambda\over 2m} 
  + 3 - 8 \log 2 \right]\left({\lambda\over 16 \pi m}\right)^2 \Bigg\} .
\label{ms2}
\end{equation}

Now, we can compute the screening mass by
inserting~(\ref{mlambda}) and~(\ref{efflam}) in~(\ref{ms2}).
We find
\begin{eqnarray}
m_s^2 &=& {1\over 24} \; g^2(2 \pi T) \; T^2 \;
\Bigg\{1 - \sqrt{6}\; {g\over 4\pi}
\nonumber \\
&+& \left[4\log{g\over 4\pi\sqrt{6}}
    -1+ 11 \log 2-\gamma+2{\zeta'(-1)\over\zeta(-1)}\right]
    \left({g\over 4\pi}\right)^2\Bigg\}\, ,
\label{ms}
\end{eqnarray}
where we have set $\mu = 2 \pi T$. Note that the dependence
of $m^2$ on $\Lambda$ in~(\ref{mlambda}) has been cancelled
by the ultraviolet cutoff introduced to regularize the
effective theory.  The term of order $g^3$ in~(\ref{ms}) was
first computed by Dolan and Jackiw~\cite{Dolan}. The
correction of order $g^4$ was obtained by Braaten and
Nieto~\cite{eft}.

It is worth stressing the distinction between $m_s$ and
$m(\Lambda)$. The screening mass contains contributions from
both the scales $T$ and $gT$. In the full theory, it may be
obtained by resumming the infrared divergent diagrams
containing self-energy insertions. On the other hand,
$m(\Lambda)$ only contains contributions from the scale $T$.
By matching~(\ref{msone}) and~(\ref{mstwo}) we see that it can
be obtained by evaluating the screening mass in the full
theory by using the strict perturbation expansion.  We can
therefore think of it as the contribution to $m_s$ from the
scale $T$.

The technique we have just applied to a massless scalar
field with $\Phi^4$ self-interaction remains essentially
unchanged when applied to QCD. In the following sections, we
will see how it is applied to computing the free energy of
QCD.

\section{Separation of Scales in QCD}\label{separation}

The partition function of QCD is
\begin{equation}\label{aqcd}
  {\cal Z} =
    \int{\cal D}A_\mu({\bf x},\tau){\cal D}q{\cal D}\overline{q}
    \exp\left\{
      -\int_0^\beta d\tau\int d{\bf x}\, {\cal L}_{\rm QCD}
  \right\}\, ,
\end{equation}
where the lagrangian is
\begin{equation}\label{Lqcd}
  {\cal L}_{\rm QCD} = {1\over 4}G_{\mu\nu}^a G_{\mu\nu}^a +
    \overline{q}\gamma_\mu D_{\mu} q  \, ,
\end{equation}
$G^a_{\mu \nu} = \partial_\mu A^a_\nu - \partial_\nu A^a_\mu
+ g f^{abc} A^b_\mu A^c_\nu$ is the field strength, and $g$
is the gauge coupling constant.  All the quark fields have
been assembled into the multi-component spinor $q$, and the
gauge-covariant derivative acting on this spinor is $D_\mu =
\partial_\mu + i g A_\mu^a T^a$.  Quarks are considered to
be massless; therefore, our description is accurate provided
that $T$ is much greater than the masses of the quarks
considered ($m_q$). Corrections from the nonzero quark
masses are suppressed by powers of $m_q/T$. Quarks with
masses $M_q$ much greater than $T$ can also be neglected as
they give corrections suppressed by powers of $T/M_q$.

We will express our calculations in terms of the
group-theory factors $C_A$, $C_F$, and $T_F$ defined by
\begin{eqnarray}
f^{abc} f^{abd} &=& C_A \delta^{cd} \;,
\\
\left( T^a T^a \right)_{ij} &=& C_F \delta_{ij} \;,
\\
{\rm tr} \left( T^a T^b \right) &=& T_F \delta^{ab} \;.
\end{eqnarray}
For an $SU(N_c)$ gauge theory with $n_f$ quarks in the
fundamental representation with masses much smaller that
$T$, these factors are $C_A = N_c$, $C_F = (N_c^2 - 1)/(2
N_c)$, and $T_F = n_f/2$.  The dimensions of the adjoint
representation and the fermion representation are $d_A =
N_c^2 - 1$ and $d_F = N_c n_f$, respectively.
 
Now we consider the effective theory that results from
integrating out the nonstatic modes, which is called
Electrostatic QCD (EQCD). Such a theory is only made out of
the static bosonic modes. The free energy density of QCD,
$F=-T\log{\cal Z}/V$ can be expressed in terms of the
effective theory as
\begin{equation}\label{FEQCD}
  F = T \left( f_E - {\log {\cal Z}_{\rm EQCD}\over V} \right)\, ,
\end{equation}
where
\begin{equation}
  {\cal Z}_{\rm EQCD} = \int{\cal D}A_0({\bf x}){\cal D}A_i({\bf x})
  \exp\left\{
      -\int d{\bf x}\, {\cal L}_{\rm EQCD}
  \right\}\, .
\end{equation}
and $f_E$ is the coefficient of the unit operator that is omitted from the
lagrangian. One can interpret $f_E T$ as the contribution to the
free energy from the scale $T$. 

The lagrangian is
\begin{equation}\label{LEQCD}
  {\cal L}_{\rm EQCD} = {1\over 4} G_{ij}^a G_{ij}^a +
    {1\over 2} (D_i A_0^a)(D_i A_0^a) + {1\over 2} m_E^2 A_0^a A_0^a +
    {1\over 8} \lambda_E (A_0^a A_0^a)^2 + \delta{\cal L}_{\rm EQCD} \, ,
\end{equation}
where $G^a_{ij} = \partial_i A^a_j - \partial_j A^a_i
        + g_E f^{abc} A^b_i A^c_j$
is the magnetostatic field strength with effective
gauge coupling constant $g_E$.
$\delta{\cal L}_{\rm EQCD}$ represents an infinite series of
non-renormalizable terms. Note that the effective field
theory does not have an effective quark field because
fermionic modes are integrated out completely; all the
effects of the fermions are incorporated into the effective
parameters.

Figures~\ref{loeqcd}$(a)$, $(b)$, $(c)$, and $(d)$ show
diagrams that contribute at leading order to the parameters
of EQCD: $f_E$, $m_E^2$, $g_E^2$, and $\lambda_E$
respectively.  Counting the powers of $g$ in the diagrams
and using dimensional analysis to determine the powers of
$T$, we find that the magnitudes of the parameters are
\begin{eqnarray}
  f_E       & \sim &     T^3 \\
  m_E^2     & \sim & g^2 T^2 \\
  g_E^2     & \sim & g^2 T   \label{pc}\\
  \lambda_E & \sim & g^4 T   \, .
\end{eqnarray}

The electrostatic field $A_0$ has a mass of order $gT$ while
the magnetostatic fields remain massless. Therefore, we can
go further in separating the different scales of QCD at high
temperature by integrating out $A_0$. We obtain an
effective theory of EQCD which is called magnetostatic QCD
(MQCD). The free energy density of QCD can be written
\begin{equation}
  F = T\left( f_E + f_M
    - {\log {\cal Z}_{\rm MQCD}\over V} \right)\, ,
\end{equation}
where
\begin{equation}
  {\cal Z}_{\rm MQCD} = \int{\cal D}A_i({\bf x})
  \exp\left\{
      -\int d{\bf x}\, {\cal L}_{\rm MQCD}
  \right\}\, .
\end{equation}
and
\begin{equation}
  {\cal L}_{\rm MQCD} = {1\over 4} G_{ij}^a G_{ij}^a +
    \delta{\cal L}_{\rm MQCD}\, .
\end{equation}
The lagrangian is only made out of magnetostatic fields
with gauge coupling constant $g_M$; $\delta{\cal L}_{\rm
MQCD}$ represents an infinite series of non-renormalizable
terms.

Similarly as we did in the case of EQCD, we can identify the
order of the leading contribution to the MQCD
parameters. Figures~\ref{lomqcd}~$(a)$ and~$(b)$ show
diagrams in EQCD that contribute at leading order to $f_M$
and $g^2_M$ respectively. Counting the powers of $g_E$ in
the diagrams and using dimensional analysis to determine the
powers of $m_E$, we find
\begin{eqnarray}
  f_M       & \sim & m_E^3 \sim g^3 T^3 \\
  g_M^2     & \sim & g_E^2 \sim g^2 T   \, .
\end{eqnarray}
We can also see the order at which $\lambda_E$ contributes
by identifying its leading contribution to $f_M$; it is
shown in Figure~\ref{lomqcd} $(c)$ and turns out to be of
order $\lambda_E m_E^2 \sim g^6 T^3$. We see that it
contributes to the free energy at order $g^6$; thus, if we
are interested in the free energy at lower order, we can
ignore $\lambda_E$. Similarly the non-renormalizable terms
of EQCD can also be omitted.

The free energy of QCD can be written as
\begin{equation}\label{fall}
  F = T\left[ f_E(T,g;\Lambda_E) +
              f_M(m^2_E, g_E, \lambda_E,
                \ldots;\Lambda_E, \Lambda_M) +
              f_G(g_M, \ldots; \Lambda_M)
      \right]\, ,
\end{equation}
where we have defined
\begin{equation}
  f_G \equiv - {\log {\cal Z}_{\rm MQCD}\over V} \, .
\end{equation}
The factorization scales $\Lambda_E$ and $\Lambda_M$
separate the scales $T$, $gT$, and $g^2 T$. $F$ is written
as the sum of three terms each of them depending only on the
parameters of the corresponding theory; their divergences
are regulated by the factorization scale parameters.

We have already seen that the leading contributions to $f_E$
and $f_M$ are of order $T^3$ and $m_E^3\sim g^3 T^3$
respectively. The leading contribution to $f_G$ can be
obtaining by realizing that the only parameter with
dimensions involved in MQCD is $g_M$ and consequently the
leading contribution is of order $g_M^6\sim g^6 T^3$. Since
we are interested in computing the free energy of QCD up to
order $g^5$ we can ignore the contribution from $f_G$; then,
$f_E$ and $f_M$ are all that we need. We conclude that the
free energy of QCD up to order $g^5$ is given by
\begin{equation}
  F = T\left[ f_E(T,g;\Lambda_E) +
              f_M(m^2_E, g_E;\Lambda_E)
      \right]\, .
\end{equation}
Therefore, we have to determine $f_E$, $m_E^2$, and $g_E$ by
matching full QCD and EQCD and then calculate $f_M$ using
EQCD.

\section{The parameters of EQCD}
\label{parameters}

In this section we will compute the parameters that define
EQCD: $f_E$, $m_E^2$, and $g_E$. They are all computed up to
the order required to reach our goal of evaluating the free
energy of QCD at order $g^5$.

\subsection{Evaluation of $g_E^2$}

In this section, we will compute $g_E$ explicitly by
comparing the actions of QCD and EQCD. To simplify the
notation we only consider the Yang-Mills part of the
lagrangian~(\ref{Lqcd}); then, the action of pure QCD is
\begin{equation}
  S_{\rm QCD} = {1\over 4} \int_0^{1/T} d\tau
   \int d^3x\; G^a_{\mu\nu} G^a_{\mu\nu} \, ,
\end{equation}
where $G^a_{\mu \nu} = \partial_\mu A^a_\nu - \partial_\nu
A^a_\mu + g f^{abc} A^b_\mu A^c_\nu$. Now, we 
use the expansion of $A_\mu^a$ in terms of its Fourier modes,
\begin{equation}
  A_\mu^a({\bf x},\tau) = 
    T\sum_{n} (A_\mu^a)_n({\bf x})\; e^{i\omega_n\tau} \, ,
\end{equation}
and write $S_{\rm QCD}$ as the sum of two terms; the first of
them only depends on the static modes of the fields and the
second on the nonstatic modes. After integrating over
$\tau$ and rescaling $(A^b_i)_0\rightarrow (A^b_i)_0/gT$, we get
\begin{equation}\label{sqcd}
  S_{\rm QCD} = {1\over 4 g^2 T}\int d^3x\;
    G^a_{ij} G^a_{ij} + \widetilde{S}_{\rm QCD}[A_0,A_n]\, ,
\end{equation}
where $G^a_{ij} = \partial_i (A^a_j)_0 - \partial_j
(A^a_i)_0 + f^{abc} (A^b_i)_0 (A^c_j)_0$. Since the first
term does not depend on the nonstatic modes, it does not
change after integrating over the nonstatic modes. As a
consequence, it can be compared directly with the action of
the effective theory obtained from the
lagrangian~(\ref{LEQCD}). After the rescaling $A_i^a
\rightarrow (A_i^a)_0/g_E$, the action for the effective
theory is
\begin{equation}\label{seqcd}
  S_{\rm EQCD} = {1\over 4 g_E^2}\int d^3x\;
  G^a_{ij} G^a_{ij} \, .
\end{equation}
Comparing~(\ref{sqcd}) and~(\ref{seqcd}), we conclude that
\begin{equation}
  g_E^2 = g^2 T  \, .
\end{equation}
We have ignored the integration of $\widetilde{S}_{\rm
QCD}$ over the nonstatic fields $(A_i^a)_n$ because we only
want to obtain $g_E$ at leading order.  This integration
contributes to $g_E$ starting at next-to-leading-order
and generates the contributions to the other parameters of
the effective lagrangian~(\ref{LEQCD}) such as $m_E^2$,
$\lambda_E$, etc.

\subsection{Evaluation of $m_E$}\label{eme}

The effective mass $m_E$ is the contribution to the electric
screening mass from the momentum scale of order $T$.  In
QCD, the electric screening mass $m_{\rm el}$
describes the asymptotic behavior of the potential between
two color charges as in~(\ref{potmel-2}). The
electric screening mass in QCD is sensitive to magnetostatic
screening effects~\cite{nadkarni-1} and therefore requires a
nonperturbative definition~\cite{ay}. However, if one
imposes an infrared cutoff that removes those magnetostatic
effects~\cite{polyakov}, one can define a perturbative
electric screening mass in terms of the location
of the pole in the gluon propagator as in~(\ref{polesm}). 
Although the gluon self-energy is gauge dependent,
the location of the pole of the propagator is gauge
invariant~\cite{kkr}. Therefore, if we use a gauge-invariant
infrared cutoff like dimensional regularization, then the
perturbative electric screening mass is gauge-invariant.

In full QCD with an infrared cutoff, the perturbative
electric screening mass $m_{\rm el}$ is the solution to the
equation
\begin{equation}
k^2 \;+\; \Pi(k^2) \;=\; 0
\qquad \mbox{at $k^2=-m_{\rm el}^2$},
\label{msdef}
\end{equation}
where $\Pi(k^2)$ is obtained from the $\mu=\nu=0$ component
of the gluon self-energy tensor evaluated at $k_0=0$:
$\Pi^{ab}_{00}(k_0=0,{\bf k}) = \Pi(k^2) \delta^{ab}$.  In
EQCD with an infrared cutoff, the perturbative electric
screening mass $m_{\rm el}$ gives the location of the pole
in the propagator for the field $A_0^a({\bf x})$.  Denoting
the self-energy function by $\Pi_E(k^2) \delta^{ab}$,
$m_{\rm el}$ is the solution to
\begin{equation}
k^2 \;+\; m_E^2 \;+\; \Pi_E(k^2) \;=\;0
\qquad \mbox{at $k^2=-m_{\rm el}^2$}.
\label{msdefeff}
\end{equation}
By matching the expressions for $m_{\rm el}$ obtained by
solving (\ref{msdef}) and (\ref{msdefeff}), we can determine
the parameter $m_E^2$.

We calculate the perturbative electric mass $m_{\rm el}$ in
the full theory using a strict perturbation expansion in
$g^2$ and using dimensional regularization with $3-2
\epsilon$ spatial dimensions to cut off both infrared and
ultraviolet divergences. Expanding the function
$\Pi^{(1)}(-m_s^2)$ on the right side of~(\ref{mst2}) in
powers of $m_s^2$ and using the fact that the solution at
lowest order is $m_s^2=\Pi^{(1)}(0)$, the resulting
expression for the perturbative electric screening mass to
next-to-leading order in $g^2$ is
\begin{equation}
m_{\rm el}^2
\;\approx\; \Pi^{(1)}(0)
\;-\; \Pi^{(1)}(0) {d \Pi^{(1)} \over d k^2}(0)
\;+\; \Pi^{(2)}(0) \;.
\label{melsq}
\end{equation}
The one-loop diagrams that contribute to $\Pi^{(1)}(k^2)$
are shown in Fig.~\ref{one} and two-loop
diagrams that contribute to $\Pi^{(1)}(k^2)$ are shown in
Fig.~\ref{two}; their analytical evaluation is
detailed in~\cite{bn5}.
We find that the strict perturbation expansion for $m_{\rm el}^2$
to order $g^4$ is
\begin{eqnarray}
m_{\rm el}^2 &\approx&
{1 \over 3} g^2(\Lambda) T^2
\Bigg\{
C_A + T_F
\nonumber \\
&& \;+\; \epsilon \left[
C_A \left( 2 {\zeta'(-1) \over \zeta(-1)}
        + 2 \log {\Lambda \over 4 \pi T} \right)\right.
\nonumber \\
&& \left. 
   \;+\; T_F \left( 1 - 2 \log 2 + 2 {\zeta'(-1) \over \zeta(-1)}
        + 2 \log {\Lambda \over 4 \pi T} \right)
\right]
\nonumber \\
&& \;+\; \left[
C_A^2 \left( {5 \over 3} + {22 \over 3} \gamma
        + {22 \over 3} \log {\Lambda \over 4 \pi T} \right)
\right.
\nonumber \\
&& \;+\; C_A T_F \left( 3 - {16 \over 3} \log 2 + {14 \over 3} \gamma
        + {14 \over 3} \log{\Lambda \over 4 \pi T} \right)
\nonumber \\
&& \left.
\;+\; T_F^2 \left( {4 \over 3} - {16 \over 3} \log 2 - {8 \over 3} \gamma
        - {8 \over 3} \log{\Lambda \over 4 \pi T} \right)
\;-\; 6 C_F T_F \right] \left({g\over 4\pi}\right)^2
\Bigg\} \;.
\label{melpert}
\end{eqnarray}
In the order $g^2$ term, we have kept terms of order
$\epsilon$ for later use.  

The expression (\ref{melpert}) for $m_{\rm el}^2$ is an
expansion in powers of $g^2$.  It does not include a $g^3$
term, in contrast to the expression for $m_{\rm el}^2$ that
correctly incorporates the effects of the screening of
electrostatic gluons \cite{rebhan}.  This $g^3$ term arises
because the $g^4$ correction includes a linear infrared
divergence that is cut off at the scale $gT$.  Since we have
used dimensional regularization as an infrared cutoff, power
infrared divergences such as this linear divergence have
been set equal to zero.

In order to match with the expression (\ref{melpert}), we
have to calculate the perturbative electric screening mass
$m_{\rm el}$ in EQCD using the strict expansion in $g^2$.
Since $m_E^2$ is treated as a perturbation parameter of
order $g^2$, the only scale in the self-energy function
$\Pi_E(k^2)$ is $k^2$.  After Taylor expanding in powers of
$k^2$, there is no scale in the dimensionally regularized
integrals; therefore, they all vanish.  The solution to the
equation (\ref{msdefeff}) for the perturbative electric
screening mass is therefore trivial:
\begin{equation}
m_{\rm el}^2 \;\approx\; m_E^2 .
\label{msperteff}
\end{equation}

Now, we compare (\ref{melpert}) and (\ref{msperteff}) and
take the limit $\epsilon \to 0$. Note that the expression
(\ref{melpert}) depends on $\Lambda$ explicitly through
logarithms of $\Lambda/4 \pi T$ and implicitly through the
coupling constant $g^2(\Lambda)$.  We shift the scale of the
coupling constant from $\Lambda$ to an arbitrary scale $\mu$
by using the renormalization group equation for the running
coupling constant
\begin{equation}
g^2(\Lambda) \;=\; g^2(\mu)
\left[ 1 \;+\; {2(11 C_A - 4 T_F) \over 3} \left({g\over 4\pi}\right)^2
                \log {\mu \over \Lambda} \right] .
\label{glam}
\end{equation}
After making this shift in the scale of the coupling
constant, the remaining $\Lambda$ can be identified with the
factorization scale $\Lambda_E$ that separates the scales
$T$ and $gT$. We conclude that the parameter $m_E^2$ is
given by
\begin{eqnarray}
m_E^2
&=&  {1 \over 3} \; g^2(\mu) \; T^2
\Bigg\{ C_A + T_F
\nonumber \\
&& \;+\; \left[
C_A^2 \left( {5 \over 3} + {22 \over 3} \gamma
        + {22 \over 3} \log {\mu \over 4 \pi T} \right)
\right.
\nonumber \\
&& \;+\; C_A T_F \left( 3 - {16 \over 3} \log 2 + {14 \over 3} \gamma
        + {14 \over 3} \log{\mu \over 4 \pi T} \right)
\nonumber \\
&& \left.
\;+\; T_F^2 \left( {4 \over 3}- {16 \over 3} \log 2 - {8 \over 3} \gamma
        - {8 \over 3} \log{\mu \over 4 \pi T} \right)
\;-\; 6 C_F T_F \right]
         \left({g\over 4\pi}\right)^2
\Bigg\} \;.
\label{mE}
\end{eqnarray}
At this order in $g^2$, there is no dependence on the
factorization scale $\Lambda_E$.

\subsection{Evaluation of $f_E$}

In this subsection, we calculate the coefficient of the unit
operator $f_E$ to next-to-next-to-leading order in
$g^2$. The physical interpretation of $f_E$ is that $f_E T$
is the contribution to the free energy from large momenta of
order $T$.  The parameter $f_E$ is determined by calculating
the free energy as a strict perturbation in $g^2$ in both
full QCD and EQCD, and matching the two results.

In the full theory, the free energy has a diagrammatic
expansion that begins with the one-loop, two-loop and
three-loop diagrams shown in Figs.~\ref{three}, \ref{four},
and~\ref{five}.  Evaluating these diagrams in Feynman gauge,
we obtain after renormalization of the coupling constant
(details can be found in~\cite{bn5})
\begin{eqnarray}
F &\approx& - {\pi^2 d_A\over 9} T^4
\Bigg\{
{1 \over 5} \;+\; {7 \over 20} {d_f \over d_A}
\;-\; \left( C_A + {5 \over 2} T_F \right) \left({g\over 4\pi}\right)^2
\nonumber \\
&& \;+\;
\left[ C_A^2
\left( {12 \over \epsilon}
        + {194 \over 3} \log{\Lambda \over 4\pi T}
        + {116 \over 5}
        + 4 \gamma + {220 \over 3} {\zeta'(-1) \over \zeta(-1)}
        - {38 \over 3} {\zeta'(-3) \over \zeta(-3)} \right)
\right.
\nonumber \\
&& \left.
\;+\; C_A T_F
\left( {12 \over \epsilon}
        + {169 \over 3} \log{\Lambda \over 4\pi T}
\right.\right.
\nonumber \\
&& \left.\left.
        \qquad\qquad + {1121 \over 60} - {157 \over 5} \log 2
        + 8 \gamma + {146 \over 3} {\zeta'(-1) \over \zeta(-1)}
        - {1\over 3} {\zeta'(-3) \over \zeta(-3)} \right)
\right.
\nonumber \\
&& \left.
\;+\; T_F^2
\left( {20 \over 3} \log{\Lambda \over 4\pi T}
        + {1 \over 3} - {88 \over 5} \log 2
        + 4 \gamma + {16 \over 3} {\zeta'(-1) \over \zeta(-1)}
        - {8 \over 3} {\zeta'(-3) \over \zeta(-3)} \right)
\right.
\nonumber \\
&& \left.
\;+\; C_F T_F
\left( {105 \over 4} - 24 \log 2 \right) \right]
        \left({g\over 4\pi}\right)^4
\Bigg\} \, ,
\label{Fpert}
\end{eqnarray}
where $g=g(\Lambda)$. The symbol ``$\approx$'' is a reminder
of the strict perturbation expansion of the full theory.

In EQCD, the free energy is given by the expression
(\ref{FEQCD}).  We calculate $\log {\cal Z}_{\rm EQCD}$
using the strict perturbation expansion in which $g_E^2$ and
$m_E^2$ are treated as perturbation parameters and both
infrared and ultraviolet divergences are regularized using
dimensional regularization.  Since diagrams with massless
propagators and with no external legs vanish in dimensional
regularization, the only contribution to $\log {\cal Z}_{\rm
EQCD}$ which does not vanish comes from the counterterm
$\delta f_E$ which cancels ultraviolet divergences
proportional to the unit operator.  The resulting expression
for the free energy is simply
\begin{equation}
F \; \approx \; ( f_E + \delta f_E ) \; T \;.
\label{Feff}
\end{equation}
The counterterm can be determined by calculating the
ultraviolet divergences in $\log {\cal Z}_{\rm EQCD}$.  If
we use dimensional regularization together with a minimal
subtraction renormalization scheme in the effective theory,
then $\delta f_E$ is a polynomial in $g_E^2$, $m_E^2$, and
the other parameters in the lagrangian for EQCD.  The only
combination of parameters that has dimension 3 and is of
order $g^4$ is $g_E^2 m_E^2$.  Thus the leading term in
$\delta f_E$ is proportional to $g_E^2 m_E^2$.  The
coefficient is determined by a 2-loop calculation that is a
trivial part of the 3-loop calculation in
Section~\ref{feog5}.  The result for the counterterm is
\begin{equation}
\delta f_E \;=\;
- {d_A C_A \over 4 (4 \pi)^2} g_E^2 m_E^2 {1 \over \epsilon} \;.
\label{deltafE}
\end{equation}
When expressing this counterterm in terms of the parameters
$g$ and $T$ of the full theory, we must take into account
the fact that $m^2_E$ multiplies a pole in $\epsilon$.  Thus
in addition to expression for $m_E^2$ given in (\ref{mE}),
we must also include the terms of order $\epsilon$ which can
be extracted from~(\ref{melpert}).  The
counterterm~(\ref{deltafE}) is therefore
\begin{eqnarray}
\delta f_E \;=\;
- {\pi^2 d_A \over 9}
        \left( {g \over 4 \pi} \right)^4 T^3 \;
\left[
12 C_A^2 \left( {1 \over \epsilon} + 2 {\zeta'(-1) \over \zeta(-1)}
        + 2 \log {\Lambda_E \over 4 \pi T} \right)
\right.
\nonumber \\
\left. \;+\; 12 C_A T_F \left( {1 \over \epsilon} + 1 - 2 \log 2
        + 2 {\zeta'(-1) \over \zeta(-1)}
        + 2 \log {\Lambda_E \over 4 \pi T} \right)
\right] \;.
\label{deltaf}
\end{eqnarray}
Note that minimal subtraction in the effective theory is not
equivalent to minimal subtraction in the full theory.  In
addition to the poles in $\epsilon$ in (\ref{deltaf}), there
are finite terms that depend on the factorization scale
$\Lambda_E$.

Matching~(\ref{Fpert}) with~(\ref{Feff}) and using the
expression (\ref{deltaf}), we conclude that $f_E$ to order
$g^4$ is
\begin{eqnarray}
f_E(\Lambda_E) &=& - {\pi^2 d_A \over 9} T^3
\Bigg\{
\left( {1 \over 5} + {7 \over 20} {d_F \over d_A} \right)
\;-\; \left( C_A + {5 \over 2} T_F \right) \left({g\over 4\pi}\right)^2
\nonumber \\
&& \hspace{-.5cm} + \Bigg(
C_A^2 \left[ 48 \log{\Lambda_E \over 4 \pi T}
        - {22 \over 3} \log{\mu \over 4 \pi T}
\right.
\nonumber \\
&& \qquad
\left.
        + {116\over 5} + 4 \gamma + {148 \over 3} {\zeta'(-1) \over \zeta(-1)}
        - {38 \over 3} {\zeta'(-3) \over \zeta(-3)} \right]
\nonumber \\
&& \hspace{-.5cm} + C_A T_F
\left[ 48  \log{\Lambda_E \over 4\pi T}
        - {47 \over 3} \log{\mu \over 4\pi T}
\right.
\nonumber \\
&&  \qquad
\left.
        + {401 \over 60} - {37 \over 5} \log 2
        + 8 \gamma + {74 \over 3} {\zeta'(-1) \over \zeta(-1)}
        - {1\over 3} {\zeta'(-3) \over \zeta(-3)} \right]
\nonumber \\
&& \hspace{-.5cm} + T_F^2
\left[ {20 \over 3} \log{\mu \over 4\pi T}
        + {1 \over 3} - {88 \over 5} \log 2
        + 4 \gamma + {16 \over 3} {\zeta'(-1) \over \zeta(-1)}
        - {8 \over 3} {\zeta'(-3) \over \zeta(-3)} \right]
\nonumber \\
&& \hspace{-.5cm} + C_F T_F
\left[ {105 \over 4} - 24 \log 2 \right] \Bigg)
\left({g\over 4\pi}\right)^4
\Bigg\} \;,
\label{fE}
\end{eqnarray}
where $g=g(\mu)$ is the coupling constant in the
$\overline{\rm MS}$ renormalization scheme at the scale
$\mu$.  We have used (\ref{glam}) to shift the scale of the
running coupling constant from $\Lambda$ to an arbitrary
renormalization scale $\mu$, and we have identified the
explicit factors of $\Lambda$ that remain with the
factorization scale $\Lambda_E$.

\section{The Free Energy of QCD}
\label{fenQCD}

Once we have understood how to resolve the contributions of
the various momentum scales in thermal QCD, asymptotic
freedom guarantees us that perturbation theory will be under
control in the high temperature limit.  At sufficiently high
temperature, the running coupling constant will be small
enough that calculations to leading order in $g$ will be
accurate.  However, in most practical applications, such as
those encountered in heavy ion collisions, the temperature
is not asymptotically large, and we must worry about higher
order corrections.  The accuracy of the perturbation
expansion can only be assessed by carrying out explicit
perturbative calculations beyond leading order.  One of the
obstacles to progress in high temperature field theory has
been that the technology for perturbative calculations was
not well developed.  Only very recently have there been any
calculations to a high enough order that the running of the
coupling constant comes into play.  The simplest physical
observable that can be calculated in perturbation theory is
the free energy, which determines all the static
thermodynamic properties of the system.  The running of the
coupling constant first enters at order $g^4$.  The free
energy for gauge theories at zero temperature but large
chemical potential was calculated to order $g^4$ long ago
\cite{mclerran}.  The first such calculation at high
temperature was the free energy of a scalar field theory
with a $\phi^4$ interaction, which was calculated to order
$g^4$ by Frenkel, Saa, and Taylor in 1992 \cite{f-s-t}.  (A
technical error was later corrected by Arnold and Zhai
\cite{arnold-zhai}.)  The analogous calculations for gauge
theories were carried out in 1994.  The free energy for QED
was calculated to order $e^4$ by Corian\`o and Parwani
\cite{coriano-parwani} and the free energy for a non-Abelian
gauge theory was calculated to order $g^4$ by Arnold and
Zhai \cite{arnold-zhai}.  The calculation of Arnold and Zhai
was completely analytic, and thus represent a particularly
significant leap in calculational technology.  The
calculational frontier has since been extended to fifth
order in the coupling constant by Parwani and
Singh~\cite{parwani-singh} and by Braaten and
Nieto~\cite{eft} for $\phi^4$ theory, by
Parwani~\cite{parwani} and Andersen~\cite{andersen}
for QED, and by Kastening and Zhai~\cite{kastening-zhai} and
Braaten and Nieto~\cite{fQCD,bn5} for non-Abelian gauge
theories.  In the following subsection, the calculation of
the free energy for a non-Abelian gauge theory to order
$g^5$ based on Ref.~\cite{fQCD,bn5} is presented.  Also, the
calculations that are required to obtain the free energy to
order $g^6$ are outlined.

\subsection{QCD free energy: up to order $g^5$}
\label{feog5}

In order to reach our goal of obtaining the free energy of QCD to
order $g^5$, it remains to evaluate the coefficient of the unit
operator of MQCD, $f_M$, which is the contribution of the scale $gT$
to the free energy.

Through order $g^5$, $f_M$ is proportional to the logarithm
of the partition function for EQCD:
\begin{equation}
f_M \;=\; - {\log {\cal Z}_{\rm EQCD} \over V} \;.
\end{equation}
The lagrangian of EQCD that we are considering at this point
is
\begin{equation}
  {\cal L}_{\rm EQCD} = {1\over 4} G_{ij}^a G_{ij}^a +
    {1\over 2} (D_i A_0)^2 + {1\over 2} m_E^2 A_0^2 \;,
\end{equation}
where now the parameters $g_E$ and $m_E$ are known. In order
to calculate the contribution to $f_M$ using perturbation
theory, we must incorporate the terms in the lagrangian that
provide electrostatic screening into the free part of the
lagrangian.  The necessary screening effects are provided by
the $A^a_0 A^a_0$ term in the EQCD lagrangian.  Thus we must
include the effects of the mass parameter $m_E^2$ to all
orders, while treating all the other coupling constants of
EQCD as perturbation parameters.  The only coupling constant
that is required to obtain the free energy to order $g^5$ is
the gauge coupling constant $g_E$.

The contributions to $\log {\cal Z}_{\rm EQCD}$ of orders
$g^3$, $g^4$, and $g^5$ are given by the sum of the 1-loop,
2-loop, and 3-loop diagrams in Figs.~\ref{six}, \ref{seven},
and~\ref{eight}.  The solid, wavy, dashed lines represent
the propagators of the $A_0$ field, the $A_i$ fields, and
the associated ghosts, respectively.  We evaluate these
diagrams in Feynman gauge.  The details of this calculation
may be found in~\cite{bn5} where the methods developed by
Broadhurst~\cite{broadhurst} were used to evaluate
analytically the integrals involved by the 3-loop diagrams.
The resulting expression for the logarithm of the partition
function is
\begin{eqnarray}
f_M
\;=\;  \;-\; {d_A \over 3 (4 \pi)} m_E^3
\;+\; {d_A C_A \over 4 (4 \pi)^2}
\left( {1 \over \epsilon} + 4 \log {\Lambda \over 2 m_E} + 3 \right)
        g_E^2 m_E^2
\nonumber \\
\;+\; {d_A C_A^2 \over (4 \pi)^3}
\left( {89 \over 24} - {11 \over 6} \log 2
        + {1 \over 6} \pi^2 \right) g_E^4 m_E
\;+\; \delta f_E \;,
\label{logZeff3}
\end{eqnarray}
where $\delta f_E$ is the counterterm associated with the
unit operator of the EQCD lagrangian and $\Lambda$ is the
scale of dimensional regularization.  It can be identified
with the ultraviolet cutoff $\Lambda_E$ of EQCD.  The
ultraviolet pole in $\epsilon$ in the term proportional to
$g^2_E m^2_E$ in (\ref{logZeff3}) is cancelled by the
counterterm $\delta f_E$, which is given in (\ref{deltafE}).
Our final result is therefore
\begin{eqnarray}
f_M(\Lambda_E) \;=\; - {d_A \over 3(4 \pi)} m_E^3 \;
\Bigg\{ 1
\;+\; \left[-3 \log {\Lambda_E \over 2 m_E} - {9 \over 4} \right]
        {C_A g_E^2 \over 4 \pi m_E}
\nonumber \\
\;+\; \left[ - {89 \over 8} + {11 \over 2} \log 2  - {1\over 2} \pi^2 \right]
        \left({C_A g_E^2 \over 4 \pi m_E}\right)^2
\Bigg\}\,.
\label{fM}
\end{eqnarray}

The coefficient $f_M$ in (\ref{fM}) can be expanded in
powers of $g$ by setting $g_E^2 = g^2T$ and by substituting
the expression (\ref{mE}) for $m_E^2$.  The complete free
energy to order $g^5$ is then $F = (f_E + f_M) T$. Note that
the dependence on the arbitrary factorization scale
$\Lambda_E$ cancels between $f_E$ and $f_M$, up to
corrections that are higher order in $g$, leaving a
logarithm of $T/m_E$.  This $g^4 \log(g)$ term is associated
with the renormalization of $f_E$, and its coefficient can
be determined from the evolution equation~\cite{bn5}
\begin{equation}
  \Lambda_E {d \ \over d \Lambda_E} f_E
  \;=\; - {d_A C_A \over (4 \pi)^2} g^2_E m^2_E + O(g^6 T^3) \;.
\label{rgfE}
\end{equation}
There is no $g^5 \log(g)$ term in the perturbation expansion
for $F$, and this is a consequence of the vanishing of the
$g_E^4$ term in the evolution equation for $m_E^2$ analogous 
to~(\ref{rgfE}).

\subsection{QCD free energy: beyond order $g^5$}

In 1980, Linde~\cite{linde} pointed out a problem concerning
perturbation theory at finite temperature. Let us consider
diagrams that contribute to the free energy of QCD like the
one in Fig.~\ref{dlinde} with $L+1$ loops. Its infrared
behavior is described by an integral of the form
\begin{equation}
  I = g^{2L} T^{L+1} \int\, \prod_{i=1}^{L+1}\, d^3k_i\,
    \prod_{j=1}^{2L}\, {1\over k_j^2 + m^2}\, ,
\end{equation}
where we have inserted a mass $m$ into the propagator as
an infrared cutoff. From dimensional analysis we see that 
when
\begin{itemize}
  \item $L<3$, $I$ is infrared finite.

  \item $L=3$, $I \sim g^6 T^4 \log(T/m)$.

  \item $L>3$, $I \sim g^6 T^4 (g^2T/m)^{L-3}$.
\end{itemize}
This means that if we consider nonstatic gluons whose mass
$m$ is set by the Matsubara frequency which is of order $T$,
then $I\sim g^6 T^4 (g^2)^{L-3}$; the magnitude of the
contribution decreases by a factor $g^2$ for each loop as in
ordinary perturbation theory. If we consider electrostatic
gluons whose screening mass is of order $gT$, then $I \sim
g^6 T^4 g^{L-3}$; the magnitude of the contribution
decreases by a factor of $g$ for each additional
loop. However, if we consider magnetostatic gluons to be
screened at a scale $m$ of order $g^2T$, $I\sim g^6 T^4$;
all of the diagrams contribute to order $g^6$, no matter how
large the number of loops is.

This problem remained open for many years. In 1994,
Braaten~\cite{solution} proposed a solution of this puzzle
in the context of the effective field theory approach we are
reviewing here.  Later, Braaten and Nieto~\cite{bn5}
analyzed the diagrams that contribute to order $g^6$ from
the different scales.  The free energy is given
by~(\ref{fall})
\begin{equation}
  F = T(f_E + f_M + f_G) \, ,
\end{equation}
where $f_E$, $f_M$, and $f_G$ give the contributions to the
free energy from the scales $T$, $gT$, and $g^2T$
respectively. We proceed to outline the calculations that
would be required to obtain $F$ to an accuracy of $g^6$.
In the full theory, we have to calculate contributions to $f_E$
from 4-loop diagrams and, also, the terms up to
order $g^4$ and $\epsilon g^4$ for $g_E^2$ and
$\lambda_E$.  In EQCD, we need to calculate the 4-loop diagrams that give
the term $g_E^6$ in $f_M$ and the 1-loop diagram that gives the 
$\lambda_E m_E$ term. Also, $g_E$ has to be
computed up to order $g^4$.  Finally, there is a contribution from MQCD
which can be written in the form
\begin{equation}
  f_G = \left( a + b \log{\Lambda_M\over g_M^2} \right) g_M^6\, .
\end{equation}
The number $b$ may be calculated by evaluating the
logarithmic ultraviolet divergence in 4-loop diagrams of
MQCD, but the number $a$ requires nonperturbative
calculations. It can be calculated using lattice simulations
of pure gauge theory in three dimensions. Recently, Karsch
{\em et al.\/}~\cite{Karsch} have reported lattice
calculations that estimate the value of $a$.

The contributions to the free energy at higher order in $g$
can be analyzed in a similar way. The contributions from the
scales $T$ and $gT$ can all be obtained from diagrammatic
calculations in full QCD and EQCD.  The contributions from
the scale $g^2T$ require nonperturbative calculation in
MQCD. It is clear that one can write an expansion for $F$
in powers of $g$ at arbitrary order. In this sense, there is
no breakdown of the weak-coupling expansion at any order.

\section{Convergence of Perturbation Theory}
\label{conver}

We have calculated the free energy as a perturbation
expansion in powers of $g$ to order $g^5$
\begin{equation}
  F = ( f_E + f_M) T \, ,
\end{equation}
where $f_E$ is given by~(\ref{fE}) and $f_M$ by~(\ref{fM}).
In this section, we examine the convergence of that
perturbation expansion; the analysis is based on
Refs.~\cite{fQCD,bn5}.  For simplicity, we focus on the case
of QCD with $n_f$ flavors of quarks.  The expansion of the
free energy in powers of $\sqrt{\alpha_s}$ with
$\alpha_s=g^2/(4\pi)$ is

\begin{eqnarray}
F \;=\; - {8 \pi^2 \over 45} T^4\;
\left[ F_0
\;+\; F_2  {\alpha_s(\mu) \over \pi}
\;+\; F_3  \left( {\alpha_s(\mu) \over \pi} \right)^{3/2}
\;+\; F_4  \left( {\alpha_s \over \pi} \right)^2
\right.
\nonumber \\
\left.
\;+\; F_5  \left( {\alpha_s \over \pi} \right)^{5/2}
\;+\; O(\alpha_s^3 \log \alpha_s) \right] \;.
\label{freeg}
\end{eqnarray}
The coefficients in this expansion are
\begin{eqnarray}
F_0 &=& 1 + \textstyle{21 \over 32} n_f \;,
\\
F_2 &=& - {15 \over 4} \left( 1 + \textstyle{5 \over 12} n_f \right)\;,
\\
F_3 &=& 30 \left( 1 + \textstyle{1 \over 6} n_f \right)^{3/2} \;,
\\
F_4 &=&  237.2 + 15.97 n_f - 0.413 n_f^2
+ { 135 \over 2} \left( 1 + \textstyle{1 \over 6} n_f \right)
        \log \left[ {\alpha_s \over \pi}
                \left(1 + \textstyle{n_f \over 6} \right) \right]
\nonumber \\
&& \;-\; { 165 \over 8}
\left( 1 + \textstyle{5 \over 12} n_f \right)
\left( 1 - \textstyle{2 \over 33} n_f \right)
        \log {\mu \over 2 \pi T} \;,
\\
F_5 &=& \left( 1 + \textstyle{1 \over 6} n_f \right)^{1/2}
\Bigg[ -799.2 - 21.96 n_f - 1.926 n_f^2
\nonumber \\
&& \;+\; {495 \over 2} \left( 1 + \textstyle{1 \over 6} n_f \right)
        \left( 1 - \textstyle{2 \over 33} n_f \right)
                \log {\mu \over 2 \pi T} \Bigg] \;.
\end{eqnarray}
The coefficient $F_2$ was first given by
Shuryak~\cite{shuryak}.  The coefficient of $F_3$ was
calculated by Kapusta \cite{kapusta} and Kalashnikov and
Klimov~\cite{Kalashnikov}. The coefficient of order
$\alpha_s^2\log\alpha_s$ was obtained by
Toimela~\cite{Toimela} and the coefficient of order
$\alpha_s^2$ by Arnold and Zhai~\cite{arnold-zhai}.  The
coefficient $F_5$ has also been calculated independently by
Kastening and Zhai~\cite{kastening-zhai}.

We now ask if the expansion~(\ref{freeg}) is well-behaved.
If the series is apparently convergent, then it can
plausibly be used to evaluate the free energy.  We study the
expression~(\ref{freeg}) at different temperatures
$T>T_c\sim 200$ MeV. We choose the renormalization scale to
be $\mu = 2 \pi T$, which is the mass of the lightest
nonstatic mode. In Table~\ref{table1} we give the
contributions coming from each order in $\sqrt{\alpha_s}$,
rescaled so that the leading order term is 1. We see that
the correction of order $\alpha_s^{5/2}$ is the largest
unless the temperature $T$ is greater than 2 GeV. The term
of order $\alpha_s^{3/2}$ is smaller than the term of order
$\alpha_s$ only when the temperature is greater than about 1
TeV.
\begin{table}
  \begin{center}
  \begin{tabular}{ccc} 
    $T$ (GeV)	& $\alpha_s(2\pi T)$	& expansion for $F$ \\ \hline
    0.250	& 0.321			& $1-0.282+0.583+0.276-1.094$ \\
    0.500	& 0.239			& $1-0.210+0.374+0.010-0.524$ \\
    1		& 0.194			& $1-0.167+0.267+0.033-0.287$ \\
    2		& 0.165			& $1-0.142+0.209+0.011-0.191$ \\
    1000	& 0.074			& $1-0.0624+0.0619+0.0106-0.0242$
  \end{tabular}
  \caption{Perturbation expansion for the Free Energy $F$
	in units of $(-8\pi^2 T^4/45)F_0$ at different temperatures.}
  \label{table1}
  \end{center}
\end{table}

We can go further in our analysis provided that we have
separated the contributions from the scales $T$ and $gT$.
We proceed to study the perturbation expansion at the scale
$T$. The term $f_E$ which gives the contribution to the free
energy from the scale $T$ is given in (\ref{fE}):
\begin{eqnarray}
f_E(\Lambda_E) &=&
- {8 \pi^2 \over 45} T^3
\Bigg\{ 1 + {\textstyle{21 \over 32}} n_f
\;-\; {15 \over 4} \left(1 + {\textstyle{5 \over 12}} n_f \right)
        {\alpha_s(\mu) \over \pi}
\nonumber \\
&&
\;+\; \Bigg[ 244.9 - 17.24 n_f - 0.415 n_f^2
\nonumber \\
&&
\;-\; {165 \over 8} \left( 1 + {\textstyle{5 \over 12}} n_f \right)
        \left( 1 - {\textstyle{2 \over 33}} n_f \right)
        \log {\mu \over 2 \pi T}
\nonumber \\
&& 
\;-\; 135 \left(1 + {\textstyle{1 \over 6}} n_f \right)
                \log {\Lambda_E \over 2 \pi T} \Bigg]
        \left( {\alpha_s \over \pi} \right)^2
\;+\; O(\alpha_s^3) \Bigg\} \;.
\label{fEnum}
\end{eqnarray}
Again we choose the renormalization scale to be $\mu=2\pi
T$. We chose the factorization scale $\Lambda_E$ to be $2\pi
T$, to avoid large logarithms of $\Lambda_E/2\pi T$.
Rescaling the contributions from each order in $\alpha_s$,
we obtain the expansion of $f_E$ shown in Table~\ref{table2}.
We see that the term of order $\alpha_s$ is reasonably small
for all the values of $T$ that are given. Note that one
could choose $\Lambda_E$ so that the contribution of order
$\alpha_s^2$ cancels. Therefore, the size of the order
$\alpha_s^2$ correction is not a very good test of the
perturbation expansion.%
\begin{table}
  \begin{center}
  \begin{tabular}{ccc} 
    $T$ (GeV)	& $\alpha_s(2\pi T)$	& expansion for $f_E$ \\ \hline
    0.250	& 0.321			& $1-0.282+0.489$ \\
    0.500	& 0.239			& $1-0.210+0.271$ \\
    1		& 0.194			& $1-0.167+0.132$ \\
    2		& 0.165			& $1-0.142+0.096$  
  \end{tabular}
  \caption{Perturbation expansion for $f_E$
	in units of $(-8\pi^2 T^4/45)F_0$ at different temperatures.}
  \label{table2}
  \end{center}
\end{table}

It is interesting to study the convergence of the other
parameter of EQCD that we have computed, $m_E^2$, which is
given by~(\ref{mE}):
\begin{eqnarray}
m_E^2 &=&
4 \pi \; \alpha_s(\mu) \; T^2
\Bigg\{ 1 + \textstyle{1 \over 6} n_f
\;+\; \Bigg[ 0.612 - 0.488 n_f - 0.0428 n_f^2
\nonumber \\
&& \;+\; {11 \over 2} \left(1 + {\textstyle{1 \over 6}} n_f \right)
        \left(1 - {\textstyle{2 \over 33}} n_f \right)
        \log {\mu \over 2 \pi T} \Bigg]
        {\alpha_s \over \pi}
\;+\; O(\alpha_s^2) \Bigg\} \;.
\label{mEnum}
\end{eqnarray}
Again, setting $\mu=2\pi T$ we obtain the corrections to the
leading order result that are summarized in
Table~\ref{table3}. Here, we also see that the
next-to-leading order correction is reasonably small for all
the values of $T$. Based on these results, we conclude that the 
perturbation series for the parameters of EQCD are well-behaved 
even at temperatures as low as 250 MeV.
\begin{table}
  \begin{center}
  \begin{tabular}{ccc} 
    $T$ (GeV)	& $\alpha_s(2\pi T)$	& expansion for $m_E^2$ \\ \hline
    0.250	& 0.321			& $1-0.124$ \\
    0.500	& 0.239			& $1-0.093$ \\
    1		& 0.194			& $1-0.098$ \\
    2		& 0.165			& $1-0.083$  
  \end{tabular}
  \caption{Perturbation expansion for $m_E^2$
	in units of $4\pi\alpha_s(2\pi T)T^2(1 + n_f/6)$ 
	at different temperatures.}
  \label{table3}
  \end{center}
\end{table}

We next examine the behavior of the perturbation expansion for EQCD.
The term $f_M$ is given by~(\ref{fM}):
\begin{eqnarray}
f_M(\Lambda_E) &=& - {2 \over 3 \pi} m_E^3
\Bigg[ 1 \;-\; \left( 0.256 + {9 \over 2}
	\log{\Lambda_E \over m_E} \right)
        {g_E^2 \over 2 \pi m_E}
\nonumber \\
&&      \;-\; 27.6 \left( {g_E^2 \over 2 \pi m_E} \right)^2
        \;+\; O(g^3) \Bigg] \;.
\label{fMnum}
\end{eqnarray}
We choose the renormalization scale of EQCD $\Lambda_E$ to
be $m_E$, which is mass of the electrostatic mode.  In
Table~\ref{table4} we give the contribution at each order in
$g_E^2/m_E$ for the factor in square brackets
in~(\ref{fMnum}).  The next-to-leading order correction
could be made to vanish by a suitable choice of
$\Lambda_E$. Therefore, it is not a good test of the
convergence of the expansion. The next-to-next-to-leading
order correction is independent of $\Lambda_E$ and is
smaller than the leading order term only if the temperature
$T > 2$ GeV. Thus, the temperature for which the
perturbation series for $f_M$ is well-behaved is much higher
than that required for the parameters of EQCD to have
well-behaved perturbation series.
\begin{table}
  \begin{center}
  \begin{tabular}{ccc} 
    $T$ (GeV)	& $\alpha_s(2\pi T)$	& expansion for $f_M$ \\ \hline
    0.250	& 0.321			& $1-1.050-1.694$ \\
    0.500	& 0.239			& $1-1.047-1.262$ \\
    1		& 0.194			& $1-0.947-0.930$ \\
    2		& 0.165			& $1-0.935-0.791$ 
  \end{tabular}
  \caption{Perturbation expansion for $f_M$
	in units of $-2/(3\pi)[4\pi\alpha_s(2\pi T) T^2(1+n_f/6)]^{3/2}$ 
	at different temperatures.}
  \label{table4}
  \end{center}
\end{table}

This analysis suggests that the slow convergence of the
expansion for $F$ in powers of $\sqrt{\alpha_s}$ may be
attributed to the slow convergence of perturbation theory at
the scale $gT$.

\section{Conclusions}
\label{conclusions}

In this paper, we have reviewed effective-field-theory
methods to study the high $T$ limit of QCD. These methods
have been used to unravel the contributions to the free
energy of QCD at high temperature from the scales $T$,
$gT$, and $g^2T$. Also, the free energy has been explicitly
computed to order $g^5$ and the calculation of the $g^6$
contribution outlined.  The calculation was significantly
streamlined by using effective-field-theory methods to
reduce every step of the calculation to one that involves
only a single momentum scale.

Our explicit calculations allow us to study the convergence
of the perturbation expansion for thermal QCD.  They suggest
that perturbation theory at the scale $gT$ requires a much
smaller value of the coupling constant than perturbation
theory at the scale $T$.  At the scale $T$, perturbative
corrections are small for all temperatures $T > T_c\simeq 200$
MeV.  Of course, even if this condition is satisfied, the
perturbation expansion may break down anyway, but this is
certainly a necessary condition.  At the scale $gT$,
perturbative corrections can be small only if $T > 2$ GeV.
Thus, in order to achieve a given relative accuracy, the
temperature $T$ must be an order of magnitude larger for
perturbation theory at the scale $gT$ compared to
perturbation theory at the scale $T$.

There is a range of temperatures in which perturbation
theory at the scale $gT$ has broken down, but perturbation
theory at the scale $T$ is reasonably accurate. In this
case, one can still use perturbation theory at the scale $T$
to calculate the parameters in the EQCD lagrangian. However,
nonperturbative methods, such as lattice simulations of
EQCD, are required to calculate the effects of the smaller
momentum scales $gT$ and $g^2T$.  While one could simply
treat the entire problem nonperturbatively using lattice
simulations of full QCD, the effective-field-theory approach
provides a dramatic savings in resources for numerical
computation.  The savings come from two sources. One is the
reduction of the problem from a 4-dimensional field theory
to a 3-dimensional field theory.  The other source of
savings is that quarks are integrated out of the theory,
which reduces it to a purely bosonic problem.

\section*{Acknowledgements}
\noindent
This work was supported in part by the U.~S. Department of
Energy, Division of High Energy Physics, under Grant
DE-FG02-91-ER40690. I would like to thank E.~Braaten for
many valuable discussions and reading the manuscript
carefully.

\pagebreak

\section*{Figure Captions}

\begin{enumerate}

\item\label{tadpole}
Diagrams that contribute to the self-energy of $\phi^4$.

\item\label{poco}
Leading order contributions to the effective parameters
of ${\cal L}_{\rm eff}$ for $\phi^4$.

\item\label{oloop}
Diagrams that contribute to the self-energy up to 2-loop order
for $\phi^4$.

\item\label{ctdia}
Diagrams that contribute to the self-energy up to 2-loop order
for $\phi^4$ involving the mass counterterm.

\item\label{loeqcd}
Leading order contributions to the parameters of EQCD: $(a)$ $f_E$,
$(b)$ $m_E^2$, $(c)$ $g_E$, and $(d)$ $\lambda_E$.

\item\label{lomqcd} Leading order contributions to the
parameters of MQCD: $(a)$ $f_M$, $(b)$ $g_M^2$, and $(c)$
the first contribution to $f_M$ that involves
$\lambda_E$. Solid and wavy lines represent the propagators
of the $A_0$ field and the $A_i$ fields, respectively. The
solid blob represents the vertex associated with
$\lambda_E$.

\item\label{one}
One-loop Feynman diagrams for the gluon self-energy.  Curly lines,
solid lines, and dashed lines represent the propagators of gluons,
quarks, and ghosts, respectively.

\item\label{two}
Two-loop Feynman diagrams for the gluon self-energy.  The solid blob
represents the sum of the one-loop gluon self-energy diagrams shown
in Fig.~\ref{one}.

\item\label{three}
One-loop Feynman diagrams for the free energy of QCD.

\item\label{four}
Two-loop Feynman diagrams for the free energy of QCD.

\item\label{five}
Three-loop Feynman diagrams for the free energy of QCD.

\item\label{six}
One-loop Feynman diagram for the logarithm of the partition function
of EQCD.

\item\label{seven}
Two-loop Feynman diagram for the logarithm of the partition function
of EQCD.

\item\label{eight}
Three-loop Feynman diagrams for the logarithm of the partition function
of EQCD.

\item\label{dlinde}
Diagram that may give rise to the breakdown of perturbation theory 
in thermal QCD.

\end{enumerate}


\newpage

\begin{thebibliography}{000}
\bibitem{Appelquist-Carazzone}
  T.~Appelquist and J.~Carazzone, Phys.\ Rev.\ D {\bf 11}, 2856 (1975).

\bibitem{appelquist-pisarski}
  T.~Appelquist and R.D.~Pisarski, Phys.\ Rev.\ D {\bf 23}, 2305 (1981).

\bibitem{nadkarni-1}
  S.~Nadkarni, Phys.\ Rev.\ D {\bf 27}, 917 (1983).

\bibitem{gpy}
  D.J.~Gross, R.D.~Pisarski, and L.G.~Yaffe,
  Rev.\ Mod.\ Phys.\ {\bf 53}, 43 (1981).

\bibitem{georgi}
  H.~Georgi,
  Annu.\ Rev.\ Nucl.\ Part.\ Sci.\ {\bf 43}, 209 (1993).

\bibitem{i-r-k}
  A.~Irb\"ack~{\em et~al.\/}, Nucl.\ Phys.\ {\bf B363}, 34 (1991);
  T.~Reisz, Z.\ Phys.\ C {\bf 53}, 169 (1992);
  L.~K\"arkk\"ainen~{\em et~al.\/}, Phys.\ Let.\ B {\bf 282}, 121 (1992);
  Nucl.\ Phys.\ {\bf B395}, 733 (1993).

\bibitem{solution}
  E.~Braaten, Phys. Rev. Lett. {\bf 74}, 2164 (1995).

\bibitem{polyakov}
  E.~Braaten and A.~Nieto, Phys.\ Rev.\ Lett.\ {\bf 74}, 3530 (1995).

\bibitem{eft}
  E.~Braaten and A.~Nieto, Phys.\ Rev.\ D {\bf 51}, 6990 (1995).

\bibitem{fQCD}
  E.~Braaten and A.~Nieto, Phys.\ Rev.\ Lett.\ {\bf 76}, 1417 (1996).

\bibitem{bn5}
  E.~Braaten and A.~Nieto, Phys. Rev.\ D {\bf 53}, 3421 (1996).

\bibitem{f-k-r-s}
  K.~Farakos, K.~Kajantie, K.~Rummukainen, and M.E.~Shaposhnikov,
        Phys.\ Lett.\ B {\bf 336}, 494 (1994);
	Nucl.\ Phys.\ {\bf B425}, 67 (1994); 
		{\bf B442}, 317 (1995);
  K.~Kajantie, M.~Laine, K.~Rummukainen, and M.~Shaposhnikov,
	Nucl.\ Phys.\ {\bf B458}, 90 (1996);
		{\bf B466}, 189 (1996);
	Phys.\ Rev.\ Lett.\ {\bf 77}, 2887 (1996).

\bibitem{shap}
  M.E.~Shaposhnikov, CERN-TH/96-280 (hep-ph/9610247).

\bibitem{andersen}
 J.O. Andersen, Phys.\ Rev.\ D {\bf 53}, 7286 (1996);
 OSLO-TP-6-96 (hep-ph/9606357) to appear in Z.\ Phys.\ C.

\bibitem{ay}
P. Arnold and L.G. Yaffe, Phys.\ Rev.\ D {\bf 52}, 7208 (1995).

\bibitem{Karsch}
  F. Karsch~{\em et~al.\/}, BI-TP-96-19 (hep-lat/9605031).

\bibitem{MSSM}
  J.M.~Cline and K.~Kainulainen, CERN-TH/96-76 (hep-ph/9605235);
  M.~Losada, RU-96-25 (hep-ph/9605266);
  M.~Laine, HD-THEP-96-13 (hep-ph/9605283).

\bibitem{kapbook} 
  J.~Kapusta, 
    {\em Finite-Temperature Field Theory}, 
    Cambridge University Press (1989).

\bibitem{lebellac}
  M.~Le~Bellac,
    {\em Thermal Field Theory},
    Cambridge University Press (1996).

\bibitem{arnold-zhai}
P. Arnold and C. Zhai, Phys.\ Rev.\ D {\bf 50}, 7603 (1994);
	Phys.\ Rev.\ D {\bf 51}, 1906 (1995).

\bibitem{landsman}
  N.P.~Landsman, Nucl.\ Phys.\ {\bf B322}, 498 (1989).

\bibitem{Dolan}
  L.~Dolan and R.~Jackiw, Phys.\ Rev.\ D {\bf 9}, 3320 (1974).

\bibitem{rebhan}
A. Rebhan, Phys.\ Rev.\ D {\bf 48}, R3967 (1993);
	Nucl. Phys. {\bf B340}, 319 (1994);
E. Braaten and A. Nieto, Phys. Rev. Lett. {\bf 73}, 2402 (1994).

\bibitem{kkr}
  R.~Kobes, G.~Kunstatter, and A.~Rebhan,
  Phys.\ Rev.\ Lett.\ {\bf 64}, 2992 (1990);
  Nucl.\ Phys.\ {\bf B355}, 1 (1991).

\bibitem{mclerran}
B.A. Freedman and L.D. McLerran, Phys.\ Rev.\ D {\bf 16}, 1147 (1977);
	{\bf 16}, 1169 (1977);
V. Baluni, Phys.\ Rev.\ D {\bf 17}, 2092 (1978).

\bibitem{f-s-t}
J.~Frenkel, A.V.~Saa, and J.C.~Taylor, Phys.\ Rev.\ D {\bf 46}, 3670 (1992).

\bibitem{coriano-parwani}
C.~Corian\`o and R.R.~Parwani, Phys.\ Rev.\ Lett.\ {\bf 73}, 2398 (1994);
R.R.~Parwani and C.~Corian\`o, Nucl.\ Phys.\ {\bf B434}, 56 (1995).

\bibitem{parwani-singh}
R.~Parwani and H. Singh, Phys. Rev. D {\bf 51}, 4518 (1995).

\bibitem{parwani}
R.R.~Parwani, Phys.\ Lett.\ {\bf B334}, 420 (1994).

\bibitem{kastening-zhai}
B. Kastening and C. Zhai, Phys.\ Rev.\ D {\bf 52}, 7232 (1995).

\bibitem{broadhurst}
D.J. Broadhurst, Z. Phys.\ C {\bf 54}, 599 (1992).

\bibitem{linde}
A.D.~Linde, Rep.\ Prog.\ Phys.\ {\bf 42}, 389 (1979);
            Phys.\ Lett.\ {\bf 96B}, 289 (1980).

\bibitem{shuryak}
E. Shuryak, J.E.T.P. {\bf 47}, 212 (1978).

\bibitem{kapusta}
J.I. Kapusta, Nucl. Phys. {\bf B148}, 461 (1979).

\bibitem{Kalashnikov}
  O.K.~Kalashnikov and V.V.~Klimov, 
  Sov.\ J.\ Nucl.\ Phys.\ {\bf 33}, 847 (1981).

\bibitem{Toimela}
  T.Toimela, Phys.\ Lett.\ {\bf 124B}, 407 (1983).

\end{thebibliography}
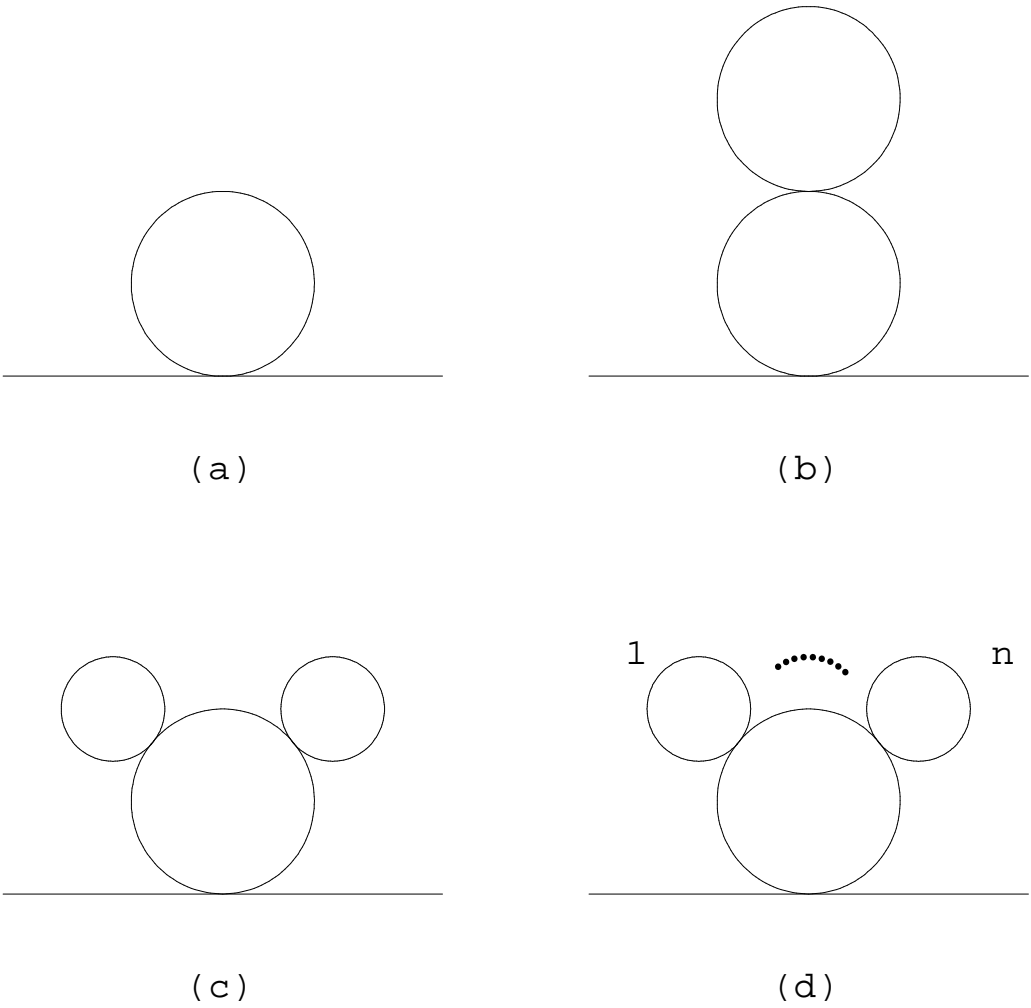
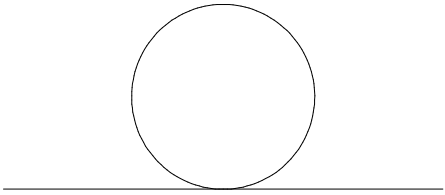
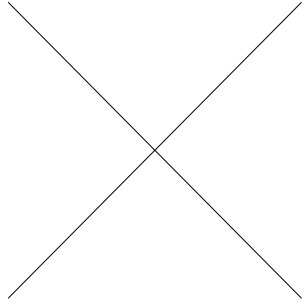
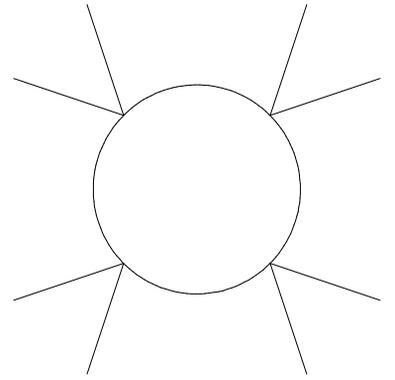
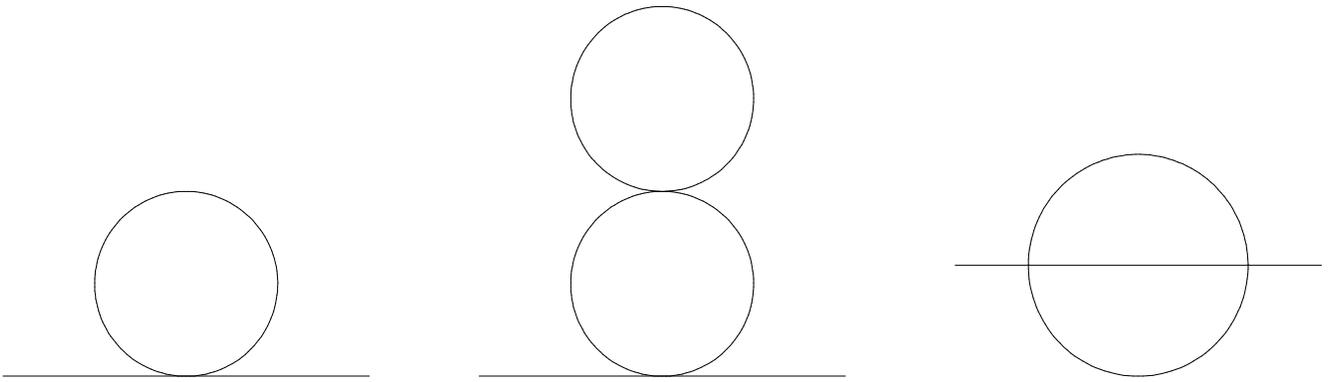
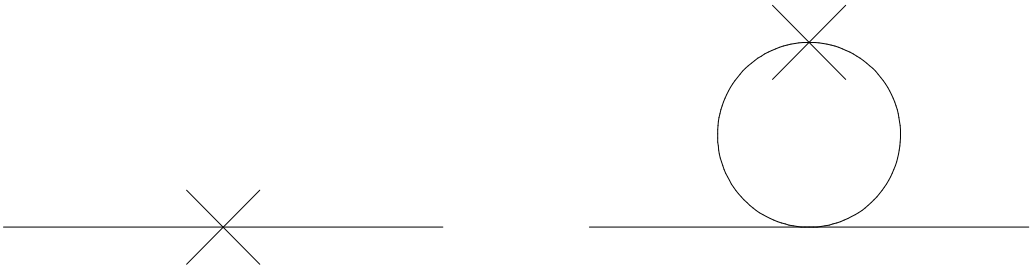
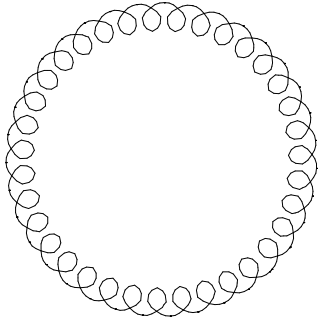
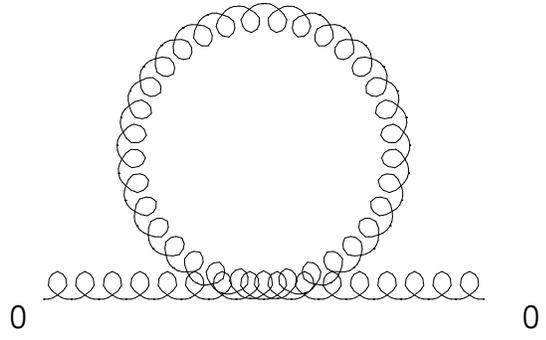
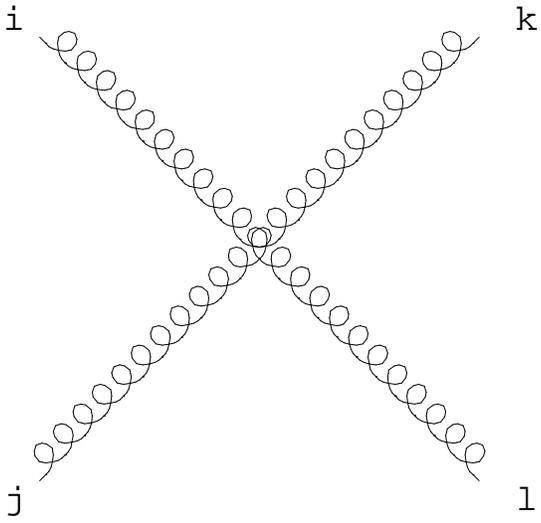
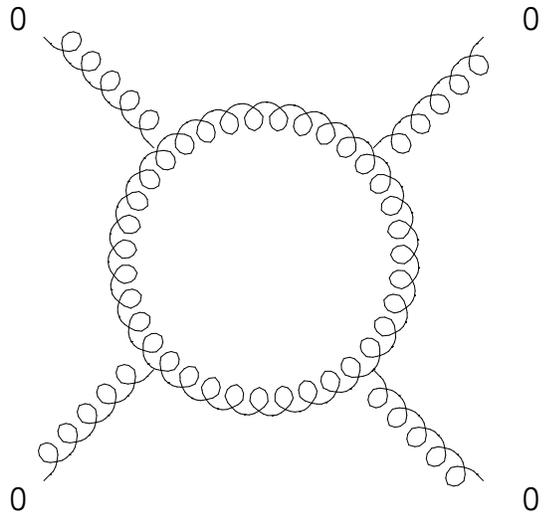
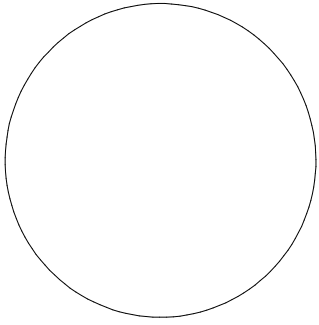
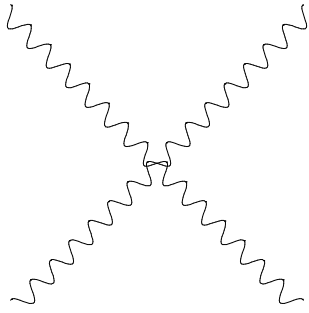
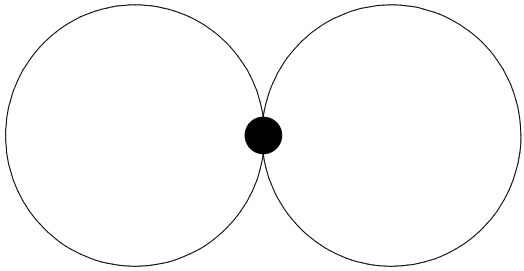
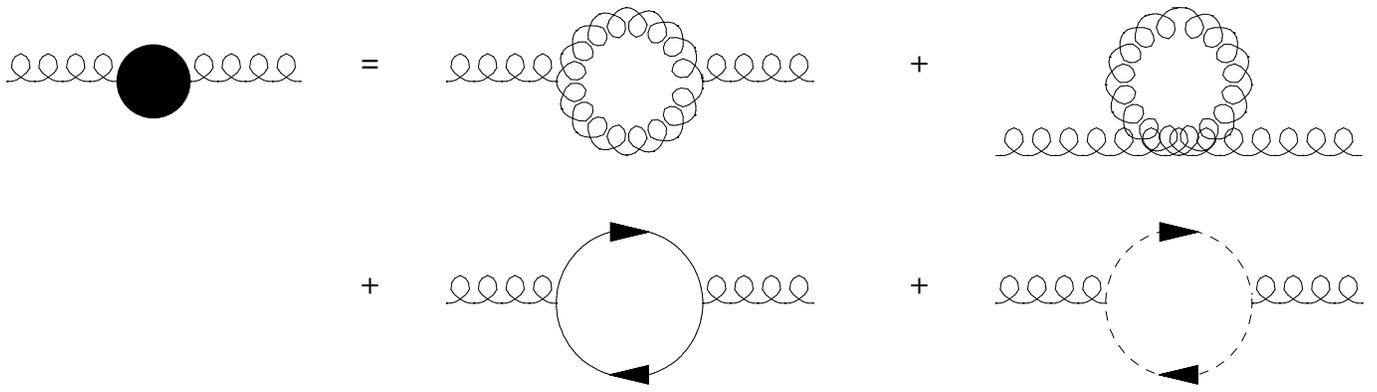
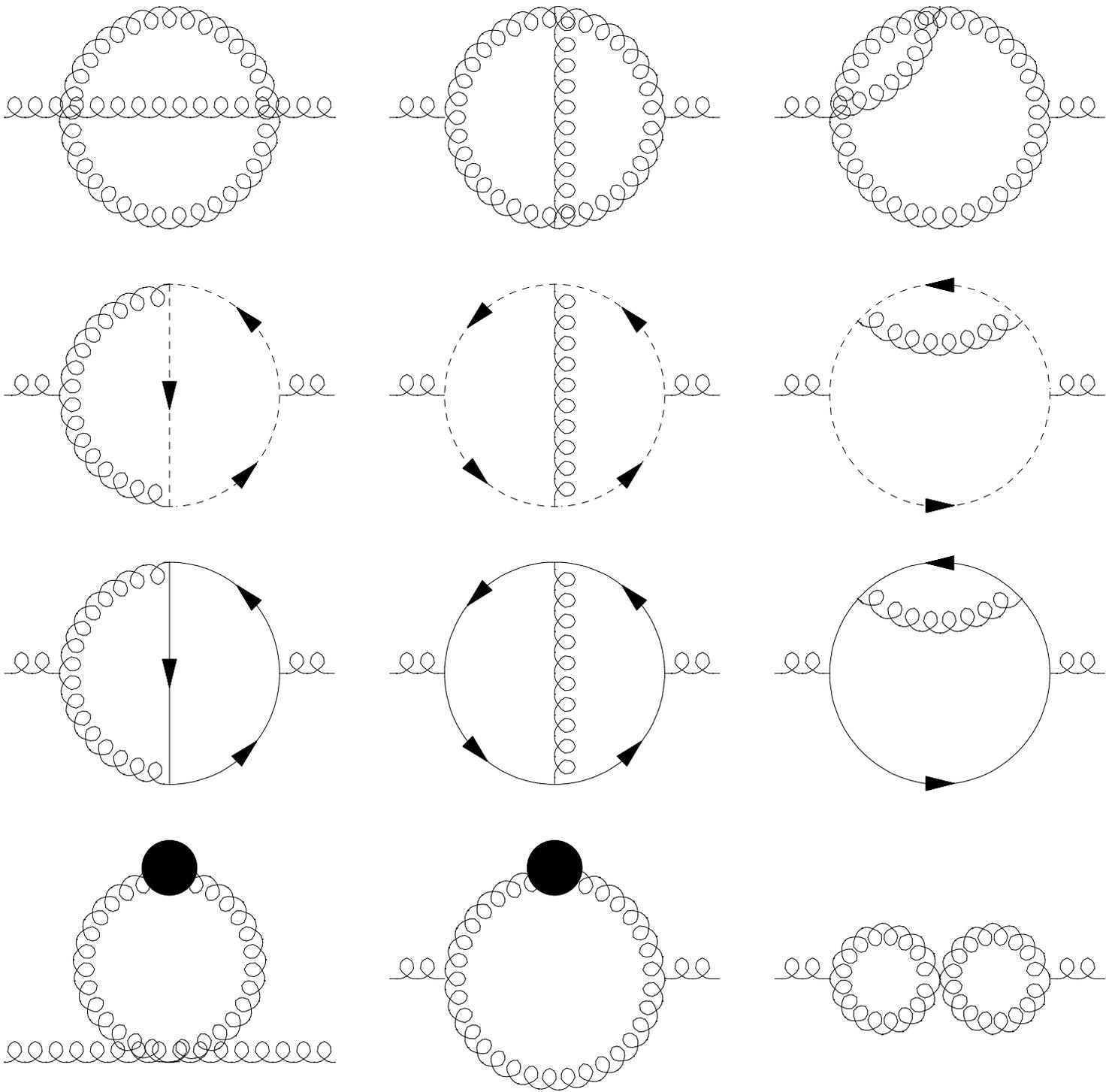
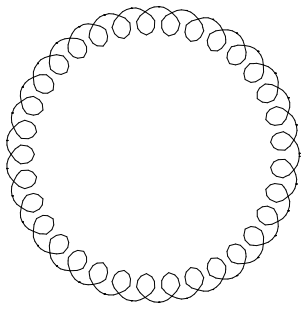
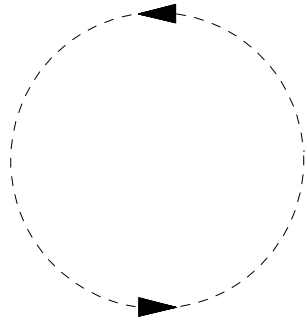
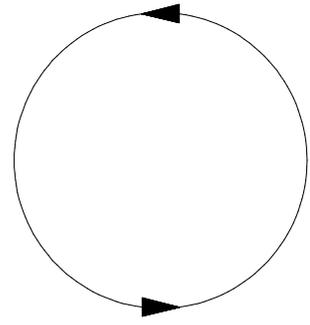
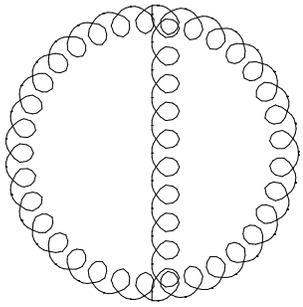
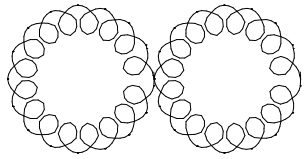
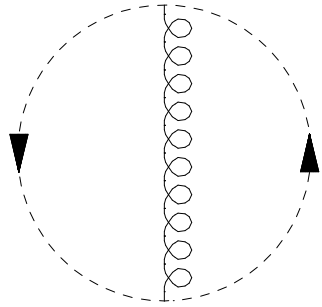
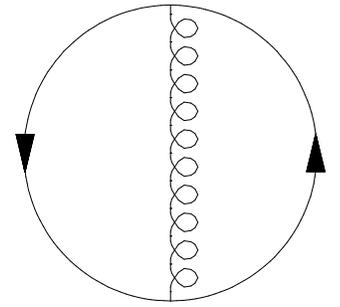
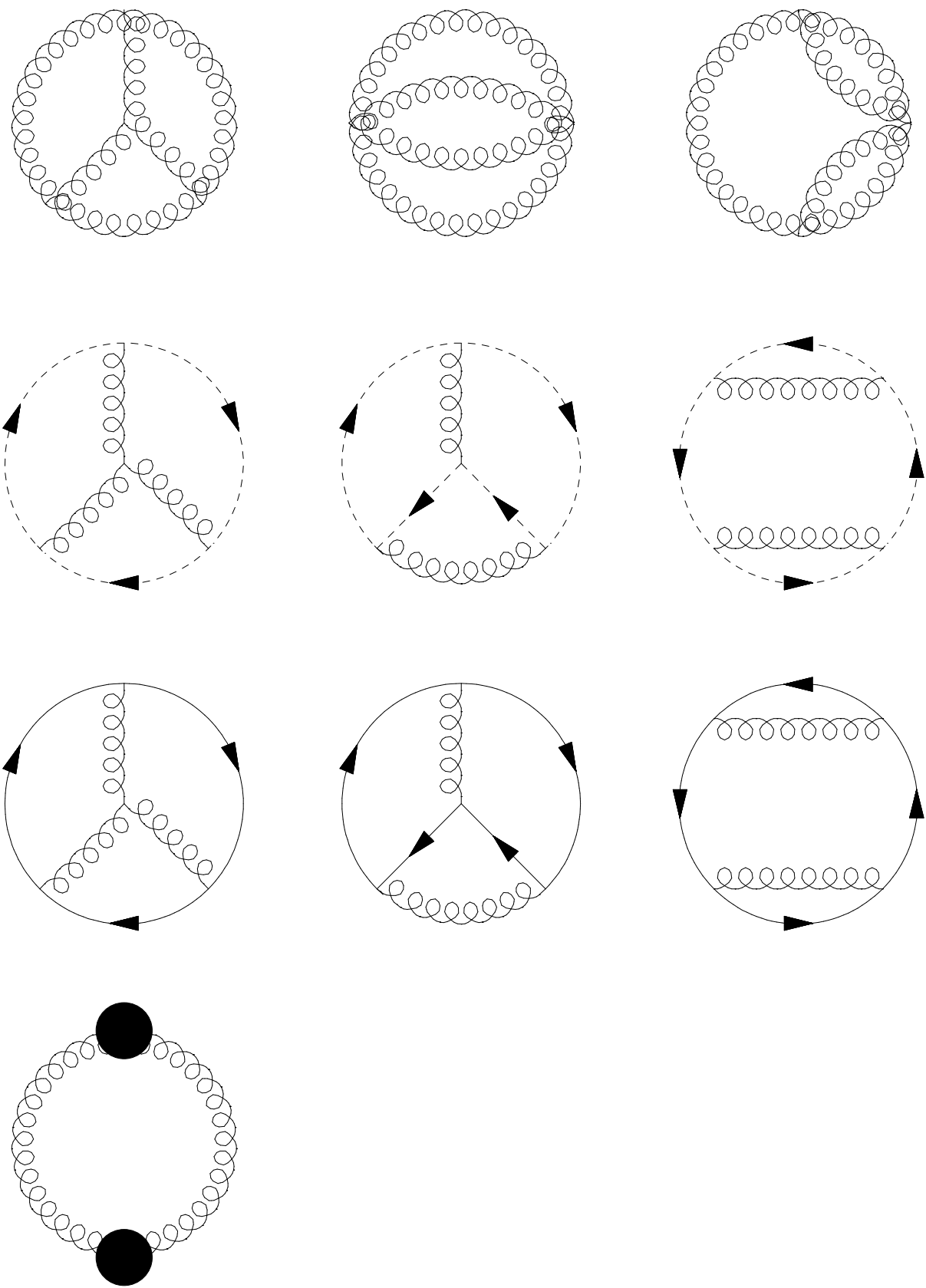
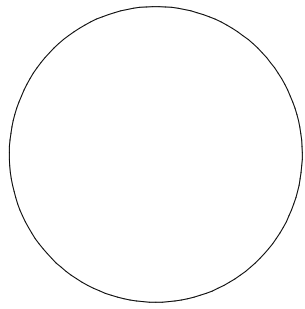
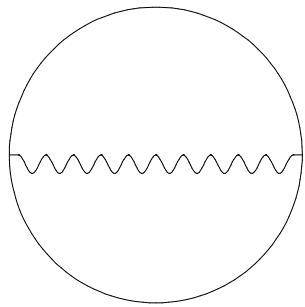
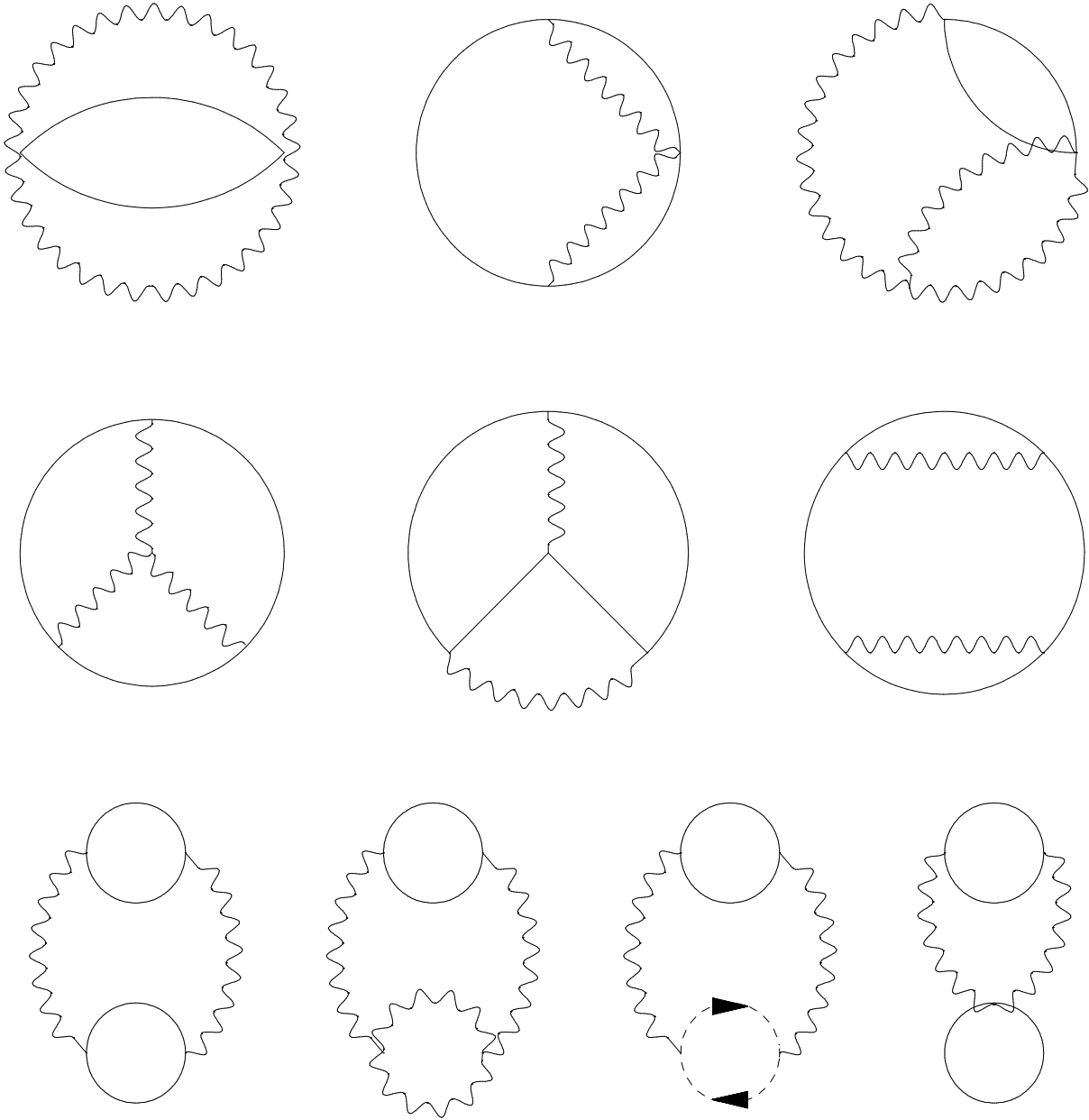
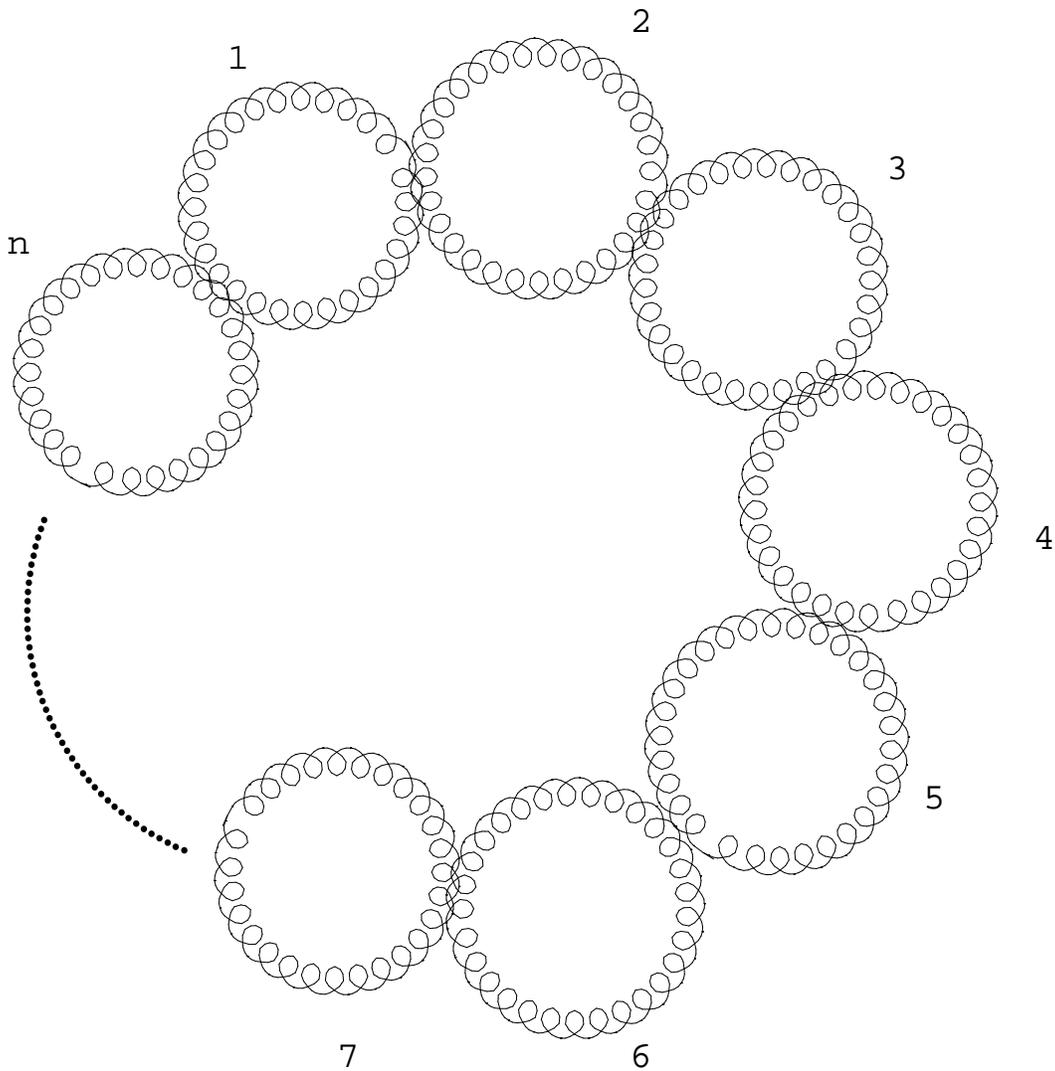
\end{document}